\documentstyle[11pt]{article}

\makeatletter
                                                                               
\def\icite{\@ifnextchar [{\@tempswatrue\@citey}{\@tempswafalse\@citey[]}}
\def\@citex[#1]#2{%
\if@filesw \immediate \write \@auxout {\string \citation {#2}}\fi 
\@tempcntb\m@ne \let\@h@ld\relax \def\@citea{}%
\@cite{%
  \@for \@citeb:=#2\do {%
    \@ifundefined {b@\@citeb}%
      {\@h@ld\@citea\@tempcntb\m@ne{\bf ?}%
      \@warning {Citation `\@citeb ' on page \thepage \space undefined}}%
      {\@tempcnta\@tempcntb \advance\@tempcnta\@ne%
      \@tempcntb\number\csname b@\@citeb \endcsname \relax%
      \ifnum\@tempcnta=\@tempcntb 
        \ifx\@h@ld\relax%
          \edef \@h@ld{\@citea\csname b@\@citeb\endcsname}%
        \else%
          \edef\@h@ld{\ifmmode{-}\else--\fi\csname b@\@citeb\endcsname}%
        \fi%
      \else
        \@h@ld\@citea\csname b@\@citeb \endcsname%
        \let\@h@ld\relax%
      \fi}%
    \def\@citea{,\penalty\@highpenalty\,}%
  }\@h@ld
}{#1}}
\def\@citey[#1]#2{%
\if@filesw \immediate \write \@auxout {\string \citation {#2}}\fi 
\@tempcntb\m@ne \let\@h@ld\relax \def\@citea{}%
\@icite{%
  \@for \@citeb:=#2\do {%
    \@ifundefined {b@\@citeb}%
      {\@h@ld\@citea\@tempcntb\m@ne{\bf ?}%
      \@warning {Citation `\@citeb ' on page \thepage \space undefined}}%
      {\@tempcnta\@tempcntb \advance\@tempcnta\@ne%
      \@tempcntb\number\csname b@\@citeb \endcsname \relax%
      \ifnum\@tempcnta=\@tempcntb 
        \ifx\@h@ld\relax%
          \edef \@h@ld{\@citea\csname b@\@citeb\endcsname}%
        \else%
          \edef\@h@ld{\ifmmode{-}\else--\fi\csname b@\@citeb\endcsname}%
        \fi%
      \else
        \@h@ld\@citea\csname b@\@citeb \endcsname%
        \let\@h@ld\relax%
      \fi}%
    \def\@citea{,\penalty\@highpenalty\,}%
  }\@h@ld
}{#1}}
\def\@cite#1#2{{$^{#1}$\if@tempswa , #2\fi }}
\def\@icite#1#2{{$#1$\if@tempswa , #2\fi }}

\gdef\@publabel{\hfil}
\gdef\@pubdate{\null}
\gdef\@pubnumber{\null}
\gdef\@author{\null}
\gdef\@title{\null}
\gdef\@abstract{\null}
\long\def\pubdate#1{\gdef\@pubdate{#1}}
\long\def\pubnumber#1{\gdef\@pubnumber{#1}}
\long\def\publabel#1{\gdef\@publabel{#1}}
\long\def\author#1{\gdef\@author{#1}}
\long\def\title#1{\gdef\@title{#1}}
\long\def\abstract#1{\gdef\@abstract{#1}}
\def\titlerelax{
\let\maketitle\relax
\let\settitleparameters\relax
\let\consolidatetitle\relax
\let\inittitlepage\relax
\let\finishtitlepage\relax
\let\titlepagecontents\relax
\let\multithanks\relax
\let\titlebaselines\relax
\let\@makepub\relax
\let\@maketitle\relax
\let\@makeauthor\relax
\let\@makeabstract\relax
\let\@maketitlenote\relax
\let\thanks\relax
\let\titlerelax\relax}

\def\titleclean
{\gdef\@titlenote{}
\gdef\@abstract{}
\gdef\@author{}
\gdef\@title{}
\gdef\@pubdate{}\gdef\@pubnumber{}\gdef\@publabel{}
\gdef\@dpublabel{}
}
\def\@makepub{\vbox to \z@{\hbox to \textwidth{\hfill
\@publabel \hfill
\llap{\parbox[t]{0.25\textwidth}{\raggedleft\@pubnumber}}}%
\vss}}
\def\@maketitle{\vskip 60pt \begin{center}
 {\LARGE \@title \par}
 \end{center}}
\def\@makeauthor{{%
\def\and{\smallskip {\normalsize \rm and\smallskip }}
\def\And{\medskip {\normalsize \rm and\\}\medskip}
\long\def\address##1{{\def\and{\\and\\}\medskip
				{\small \it \\##1\\}
}}
{\centering
 \vskip 3em
 \large \lineskip .75em
 \@author}
 \par}} 
\def\@makedate{\vskip 1.5em 
 {\raggedright \small \noindent\@pubdate \par}}
\def\@makeabstract{\vskip 1.5em
{\small 
\begin{center}
{\bf ABSTRACT\vspace{-.5em}\vspace{0pt}} 
\end{center}
\quotation \@abstract \endquotation}}
\def\maketitle{\titlepage
\let\footnotesize\small \setcounter{page}{0}
\def\thefootnote{\fnsymbol{footnote}}
\@makepub
\vfil
\@maketitle
\@makeauthor
\vfil
\@makeabstract
\@thanks
\vfil
\@makedate
\if@restonecol\twocolumn \else \eject \fi
\titlerelax \titleclean
\def\thefootnote{\alph{footnote}}
\setcounter{footnote}{0}
}

\ifcase\@ptsize
 \font\tenmsa=msam10
 \font\sevenmsa=msam7
 \font\fivemsa=msam5
 \font\tenmsb=msbm10
 \font\sevenmsb=msbm7
 \font\fivemsb=msbm5
\or
 \font\tenmsa=msam10 scaled \magstephalf
 \font\sevenmsa=msam8
 \font\fivemsa=msam6
 \font\tenmsb=msbm10 scaled \magstephalf
 \font\sevenmsb=msbm8
 \font\fivemsb=msbm6
\or
 \font\tenmsa=msam10 scaled \magstep1
 \font\sevenmsa=msam8
 \font\fivemsa=msam6
 \font\tenmsb=msbm10 scaled \magstep1
 \font\sevenmsb=msbm8
 \font\fivemsb=msbm6
\fi

\newfam\msafam
\newfam\msbfam
\textfont\msafam=\tenmsa  \scriptfont\msafam=\sevenmsa
  \scriptscriptfont\msafam=\fivemsa
\textfont\msbfam=\tenmsb  \scriptfont\msbfam=\sevenmsb
  \scriptscriptfont\msbfam=\fivemsb

\def\hexnumber@#1{\ifnum#1<10 \number#1\else
 \ifnum#1=10 A\else\ifnum#1=11 B\else\ifnum#1=12 C\else
 \ifnum#1=13 D\else\ifnum#1=14 E\else\ifnum#1=15 F\fi\fi\fi\fi\fi\fi\fi}

\def\msa@{\hexnumber@\msafam}
\def\msb@{\hexnumber@\msbfam}
\mathchardef\boxdot="2\msa@00
\mathchardef\boxplus="2\msa@01
\mathchardef\boxtimes="2\msa@02
\mathchardef\square="0\msa@03
\mathchardef\blacksquare="0\msa@04
\mathchardef\centerdot="2\msa@05
\mathchardef\lozenge="0\msa@06
\mathchardef\blacklozenge="0\msa@07
\mathchardef\circlearrowright="3\msa@08
\mathchardef\circlearrowleft="3\msa@09
\mathchardef\rightleftharpoons="3\msa@0A
\mathchardef\leftrightharpoons="3\msa@0B
\mathchardef\boxminus="2\msa@0C
\mathchardef\Vdash="3\msa@0D
\mathchardef\Vvdash="3\msa@0E
\mathchardef\vDash="3\msa@0F
\mathchardef\twoheadrightarrow="3\msa@10
\mathchardef\twoheadleftarrow="3\msa@11
\mathchardef\leftleftarrows="3\msa@12
\mathchardef\rightrightarrows="3\msa@13
\mathchardef\upuparrows="3\msa@14
\mathchardef\downdownarrows="3\msa@15
\mathchardef\upharpoonright="3\msa@16

\mathchardef\downharpoonright="3\msa@17
\mathchardef\upharpoonleft="3\msa@18
\mathchardef\downharpoonleft="3\msa@19
\mathchardef\rightarrowtail="3\msa@1A
\mathchardef\leftarrowtail="3\msa@1B
\mathchardef\leftrightarrows="3\msa@1C
\mathchardef\rightleftarrows="3\msa@1D
\mathchardef\Lsh="3\msa@1E
\mathchardef\Rsh="3\msa@1F
\mathchardef\rightsquigarrow="3\msa@20
\mathchardef\leftrightsquigarrow="3\msa@21
\mathchardef\looparrowleft="3\msa@22
\mathchardef\looparrowright="3\msa@23
\mathchardef\circeq="3\msa@24
\mathchardef\succsim="3\msa@25
\mathchardef\gtrsim="3\msa@26
\mathchardef\gtrapprox="3\msa@27
\mathchardef\multimap="3\msa@28
\mathchardef\therefore="3\msa@29
\mathchardef\because="3\msa@2A
\mathchardef\doteqdot="3\msa@2B

\mathchardef\triangleq="3\msa@2C
\mathchardef\precsim="3\msa@2D
\mathchardef\lesssim="3\msa@2E
\mathchardef\lessapprox="3\msa@2F
\mathchardef\eqslantless="3\msa@30
\mathchardef\eqslantgtr="3\msa@31
\mathchardef\curlyeqprec="3\msa@32
\mathchardef\curlyeqsucc="3\msa@33
\mathchardef\preccurlyeq="3\msa@34
\mathchardef\leqq="3\msa@35
\mathchardef\leqslant="3\msa@36
\mathchardef\lessgtr="3\msa@37
\mathchardef\backprime="0\msa@38
\mathchardef\risingdotseq="3\msa@3A
\mathchardef\fallingdotseq="3\msa@3B
\mathchardef\succcurlyeq="3\msa@3C
\mathchardef\geqq="3\msa@3D
\mathchardef\geqslant="3\msa@3E
\mathchardef\gtrless="3\msa@3F
\mathchardef\sqsubset="3\msa@40
\mathchardef\sqsupset="3\msa@41
\mathchardef\vartriangleright="3\msa@42
\mathchardef\vartriangleleft="3\msa@43
\mathchardef\trianglerighteq="3\msa@44
\mathchardef\trianglelefteq="3\msa@45
\mathchardef\bigstar="0\msa@46
\mathchardef\between="3\msa@47
\mathchardef\blacktriangledown="0\msa@48
\mathchardef\blacktriangleright="3\msa@49
\mathchardef\blacktriangleleft="3\msa@4A
\mathchardef\vartriangle="3\msa@4D
\mathchardef\blacktriangle="0\msa@4E
\mathchardef\triangledown="0\msa@4F
\mathchardef\eqcirc="3\msa@50
\mathchardef\lesseqgtr="3\msa@51
\mathchardef\gtreqless="3\msa@52
\mathchardef\lesseqqgtr="3\msa@53
\mathchardef\gtreqqless="3\msa@54
\mathchardef\Rrightarrow="3\msa@56
\mathchardef\Lleftarrow="3\msa@57
\mathchardef\veebar="2\msa@59
\mathchardef\barwedge="2\msa@5A
\mathchardef\doublebarwedge="2\msa@5B
\mathchardef\angle="0\msa@5C
\mathchardef\measuredangle="0\msa@5D
\mathchardef\sphericalangle="0\msa@5E
\mathchardef\varpropto="3\msa@5F
\mathchardef\smallsmile="3\msa@60
\mathchardef\smallfrown="3\msa@61
\mathchardef\Subset="3\msa@62
\mathchardef\Supset="3\msa@63
\mathchardef\Cup="2\msa@64

\mathchardef\Cap="2\msa@65

\mathchardef\curlywedge="2\msa@66
\mathchardef\curlyvee="2\msa@67
\mathchardef\leftthreetimes="2\msa@68
\mathchardef\rightthreetimes="2\msa@69
\mathchardef\subseteqq="3\msa@6A
\mathchardef\supseteqq="3\msa@6B
\mathchardef\bumpeq="3\msa@6C
\mathchardef\Bumpeq="3\msa@6D
\mathchardef\lll="3\msa@6E

\mathchardef\ggg="3\msa@6F

\mathchardef\circledS="0\msa@73
\mathchardef\pitchfork="3\msa@74
\mathchardef\dotplus="2\msa@75
\mathchardef\backsim="3\msa@76
\mathchardef\backsimeq="3\msa@77
\mathchardef\complement="0\msa@7B
\mathchardef\intercal="2\msa@7C
\mathchardef\circledcirc="2\msa@7D
\mathchardef\circledast="2\msa@7E
\mathchardef\circleddash="2\msa@7F
\def\ulcorner{\delimiter"4\msa@70\msa@70 }
\def\urcorner{\delimiter"5\msa@71\msa@71 }
\def\llcorner{\delimiter"4\msa@78\msa@78 }
\def\lrcorner{\delimiter"5\msa@79\msa@79 }
\def\yen{\mathhexbox\msa@55 }
\def\checkmark{\mathhexbox\msa@58 }
\def\circledR{\mathhexbox\msa@72 }
\def\maltese{\mathhexbox\msa@7A }
\mathchardef\lvertneqq="3\msb@00
\mathchardef\gvertneqq="3\msb@01
\mathchardef\nleq="3\msb@02
\mathchardef\ngeq="3\msb@03
\mathchardef\nless="3\msb@04
\mathchardef\ngtr="3\msb@05
\mathchardef\nprec="3\msb@06
\mathchardef\nsucc="3\msb@07
\mathchardef\lneqq="3\msb@08
\mathchardef\gneqq="3\msb@09
\mathchardef\nleqslant="3\msb@0A
\mathchardef\ngeqslant="3\msb@0B
\mathchardef\lneq="3\msb@0C
\mathchardef\gneq="3\msb@0D
\mathchardef\npreceq="3\msb@0E
\mathchardef\nsucceq="3\msb@0F
\mathchardef\precnsim="3\msb@10
\mathchardef\succnsim="3\msb@11
\mathchardef\lnsim="3\msb@12
\mathchardef\gnsim="3\msb@13
\mathchardef\nleqq="3\msb@14
\mathchardef\ngeqq="3\msb@15
\mathchardef\precneqq="3\msb@16
\mathchardef\succneqq="3\msb@17
\mathchardef\precnapprox="3\msb@18
\mathchardef\succnapprox="3\msb@19
\mathchardef\lnapprox="3\msb@1A
\mathchardef\gnapprox="3\msb@1B
\mathchardef\nsim="3\msb@1C
\mathchardef\napprox="3\msb@1D
\mathchardef\varsubsetneq="3\msb@20
\mathchardef\varsupsetneq="3\msb@21
\mathchardef\nsubseteqq="3\msb@22
\mathchardef\nsupseteqq="3\msb@23
\mathchardef\subsetneqq="3\msb@24
\mathchardef\supsetneqq="3\msb@25
\mathchardef\varsubsetneqq="3\msb@26
\mathchardef\varsupsetneqq="3\msb@27
\mathchardef\subsetneq="3\msb@28
\mathchardef\supsetneq="3\msb@29
\mathchardef\nsubseteq="3\msb@2A
\mathchardef\nsupseteq="3\msb@2B
\mathchardef\nparallel="3\msb@2C
\mathchardef\nmid="3\msb@2D
\mathchardef\nshortmid="3\msb@2E
\mathchardef\nshortparallel="3\msb@2F
\mathchardef\nvdash="3\msb@30
\mathchardef\nVdash="3\msb@31
\mathchardef\nvDash="3\msb@32
\mathchardef\nVDash="3\msb@33
\mathchardef\ntrianglerighteq="3\msb@34
\mathchardef\ntrianglelefteq="3\msb@35
\mathchardef\ntriangleleft="3\msb@36
\mathchardef\ntriangleright="3\msb@37
\mathchardef\nleftarrow="3\msb@38
\mathchardef\nrightarrow="3\msb@39
\mathchardef\nLeftarrow="3\msb@3A
\mathchardef\nRightarrow="3\msb@3B
\mathchardef\nLeftrightarrow="3\msb@3C
\mathchardef\nleftrightarrow="3\msb@3D
\mathchardef\divideontimes="2\msb@3E
\mathchardef\varnothing="0\msb@3F
\mathchardef\nexists="0\msb@40
\mathchardef\mho="0\msb@66
\mathchardef\thorn="0\msb@67
\mathchardef\beth="0\msb@69
\mathchardef\gimel="0\msb@6A
\mathchardef\daleth="0\msb@6B
\mathchardef\lessdot="3\msb@6C
\mathchardef\gtrdot="3\msb@6D
\mathchardef\ltimes="2\msb@6E
\mathchardef\rtimes="2\msb@6F
\mathchardef\shortmid="3\msb@70
\mathchardef\shortparallel="3\msb@71
\mathchardef\smallsetminus="2\msb@72
\mathchardef\thicksim="3\msb@73
\mathchardef\thickapprox="3\msb@74
\mathchardef\approxeq="3\msb@75
\mathchardef\succapprox="3\msb@76
\mathchardef\precapprox="3\msb@77
\mathchardef\curvearrowleft="3\msb@78
\mathchardef\curvearrowright="3\msb@79
\mathchardef\digamma="0\msb@7A
\mathchardef\varkappa="0\msb@7B
\mathchardef\hslash="0\msb@7D
\mathchardef\hbar="0\msb@7E
\mathchardef\backepsilon="3\msb@7F
\def\Bbb{\ifmmode\let\next\Bbb@\else
 \def\next{\errmessage{Use \string\Bbb\space only in math mode}}\fi\next}
\def\Bbb@#1{{\Bbb@@{#1}}}
\def\Bbb@@#1{\fam\msbfam#1}

\def\sk {{\hskip 1 cm}}
\def\bk {{\hskip 0.2 cm}}

\def\rk {{\hskip 3 cm}}
\def\com{{\hskip 0.2 cm},}
\def\pkt{{\hskip 0.2 cm}.}
\def\acknowledgements{\@startsection{section}{4}
{\z@}{-3.5ex plus -1ex minus -.2ex}{2.3ex plus .2ex}{\normalsize\bf}
{Acknowledgements}}

\def\sct {{\rm{\tt{K_2}}}}      
\newcommand{\scn}[1]{{\rm{\tt{K_{#1}}
}}}                            
\def\half {\frac{1}{2}}		  
\newcommand{\secn}[1]{{\sc Sec.}\,{\sf #1}}	  
\newcommand{\eq}[1]{{\sc Eq.}\,{\sf (#1)}}	  
\newcommand{\eqs}[1]{{\sc Eqs.}\,{\sf (#1)}}	  
\newcommand{\refoth}[1]{{\sf #1}}  	  	  
\newcommand{\eqoth}[1]{{\sf (#1)}}  	  	  

\def\bbbz {\Bbb{Z}}		  
\def\bbbzh{\Bbb{Z}_{\frac{1}{2}}} 

\def\bbbn {\Bbb{N}}   	          
\def\bbbno{\Bbb{N}_{0}}           
\def\bbbc {\Bbb{C}}

\def\bbbone {{\mathchoice {\rm 1\mskip-4mu l} {\rm 1\mskip-4mu l}
{\rm 1\mskip-4.5mu l} {\rm 1\mskip-5mu l}}}

\newtheorem{definition}{Definition}[section]
\newtheorem{theorem}[definition]{Theorem}
\newtheorem{lemma}[definition]{Lemma}
\newtheorem{proposition}[definition]{Proposition}
\newcounter{defs}[section]

\newcommand{\be}{\begin{equation}}
\newcommand{\ee}{\end{equation}}
\newcommand{\bea}{\begin{eqnarray}}
\newcommand{\eea}{\end{eqnarray}}

\newcommand{\bdf}{\stepcounter{defs}\begin{definition}}
\newcommand{\edf}{\end{definition}}
\newcommand{\bth}{\stepcounter{defs}\begin{theorem}}
\newcommand{\eth}{\end{theorem}}
\newcommand{\blm}{\stepcounter{defs}\begin{lemma}}
\newcommand{\elm}{\end{lemma}}
\newcommand{\bpr}{\stepcounter{defs}\begin{proposition}}
\newcommand{\epr}{\end{proposition}}
\newcommand{\bprf}{Proof: }
\newcommand{\eprf}{\hfill $\blacksquare$ \\}
\newcounter{pics}
\newcommand{\bpic}[4]{\begin{center}\begin{picture}(#1,#2)(#3,#4)
\refstepcounter{pics}}
\renewcommand{\thepics}{{\sf\roman{pics}}}
\newcommand{\epic}[1]{\end{picture}\\
{\small {\sc Fig.} \thepics \bk #1} \end{center}}
\newcommand{\epicspl}{\end{picture}\\		
\addtocounter{pics}{-1}\end{center}}		

\renewcommand{\thefootnote}{\rm{\alph{footnote}}}
\newcounter{tabs}
\newcommand{\btab}[1]{\refstepcounter{tabs}\begin{center}
\begin{tabular}{#1}}
\renewcommand{\thetabs}{{\sf\alph{tabs}}}
\newcommand{\etab}[1]{\end{tabular}\\[1.5ex]
{\small {\sc Tab.} \thetabs \bk #1} \end{center}}

\def\noi {\noindent}
\newcommand{\nn}{\nonumber}

\newcommand{\ket}[1]{\left| {#1} \right\rangle}	
\newcommand{\bra}[1]{\left\langle {#1} \right|}	
\newcommand{\spn}[1]{{\tt span}\{{#1}\}}	
\def\mod{{\tt mod}}				
\newcommand{\fall}[2]{(#1)^{\underline{#2}}}	
\newcommand{\der}[2]{\frac{d #1}{d #2}}	        
\newcommand{\pder}[2]{\frac{\partial #1}{
\partial #2}}	        			
\def\dif{{\rm d}}				
\def\abs{\,|\,}					
\newcommand{\trace}[1]{{\tt tr}\left( #1 
\right) }					
\def\det{{\tt det}}				
\def\without{\backslash}			

\def\pmb#1{\setbox0=\hbox{#1}%
 \kern-.025em\copy0\kern-\wd0
 \kern.05em\copy0\kern-\wd0
 \kern-.025em\raise.0433em\box0 }

\def\bPhi{{\pmb{$\Phi$}}}

\makeatother

\title{The definition of Neveu-Schwarz superconformal fields and 
uncharged superconformal transformations}
\author{Matthias D\"{o}rrzapf
\thanks{e-mail: matthias@feynman.harvard.edu} 
\address{Lyman Laboratory of Physics\\
Harvard University\\
Cambridge, MA 02138, USA }}
\pubdate{April 1998}
\pubnumber{HUTP-97/A054 \\ hep-th/9712107}

\abstract{ 
The construction of Neveu-Schwarz superconformal field
theories for any $N$ is given
via a superfield formalism. We also review some results and definitions
of superconformal manifolds and we generalise contour integration and
Taylor expansion to superconformal spaces. For arbitrary
 $N$ we define (uncharged) primary
fields and give their infinitesimal change under superconformal 
transformations. This leads us to the operator product expansion
of the stress-energy tensor with itself and with primary fields.
In this way we derive the well-known commutation relations of the Neveu-Schwarz
superconformal algebras $\scn{N}$. In this context we observe that the central
extension term disappears for $N \geq 4$ for the Neveu-Schwarz theories.
Finally, we give the global transformation rules of
primary fields under the action of the algebra generators.}

\begin{document}

\maketitle



\section{{Introduction}}


The interplay of symmetries and conservation laws is one
of the most intriguing features of physics and it can
be found in many different areas of physics. Very often,
physical systems for which one would have {\it a priori}
thought that they have nothing in common,
share in fact the same symmetry properties and thus have common 
physical properties. A recent example that illustrates this is 
local conformal symmetry, a concept which concerns various sectors
of physics. The conformal symmetry group includes both
Poincar\'e symmetry and scale invariance. The group of globally defined
conformal transformations on the Riemann sphere 
are the well-known M\"obius 
transformations, the transformations keeping angles invariant. 
In two dimensions, local conformal 
transformations
are simply the locally holomorphic functions.

One of the most fascinating facts about statistical systems is
the existence of special critical points where the systems 
become scale invariant and thus locally conformally invariant\cite{bpz}.
Sweeping to an entirely different part of physics, string theory,
we again find local conformal symmetry, here in two dimensions;
after fixing the local symmetry of the string we are left with 
a conformally invariant field theory in two flat dimensions. 
Watching out for 
conformal invariance through physics
we come across percolation systems\cite{percolation},
random walk models\cite{random} among many more
still to be discovered.

Conformal invariance in a conformal field theory of
two dimensions turns out to be particularly
interesting since the algebra of symmetry generators becomes
infinite dimensional. The algebra of generators of conformal 
transformations in two dimensions is given by 
the {\it Virsoro algebra}. This is an
infinite dimensional Lie algebra with the commutation relations
\bea
[L_{m},L_{n}] &=& (m-n) L_{m+n} + \frac{C}{12} (m^{3}-m) \delta_{m+n,0}
\com \label{eq:vir} \\
\ [L_{m},C] &=& 0 \com \rk m,n \in \bbbz \nn \pkt
\eea
For a classical theory the central extension $C$ would be trivial and
therefore \eq{\ref{eq:vir}} would represent the {\it de Witt algebra}. 
Starting with the paper of Belavin,
Polyakov and Zamolodchikov\cite{bpz}, many statistical models at their
critical points have been identified as conformally invariant 
theories\cite{abf,friedan2,huse}.
In the canonical quantisation scheme, $L_{0}$ generates time translations and 
hence represents the energy of the system. Since the energy is bounded below, 
the space of states of the physical system is confined to be a sum over
{\it highest weight representations}\footnote{There is the usual historical
confusion: what physicists call a highest weight vector is in fact a
vector of lowest weight in the Verma module.} 
of the algebra \eq{\ref{eq:vir}}. 
A highest weight 
representation of the Virasoro algebra is a representation
containing a vector $\ket{h,c}$ such that
\bea
L_{0} \ket{h,c}=h \ket{h,c}  \com \nn &
C \ket{h,c}=c \ket{h,c} \com \label{eq:hw} &
L_{n} \ket{h,c}=0 \com \sk \forall n \in \bbbn \pkt \nn
\eea
$\ket{h,c}$ is called a {\it highest weight vector} with {\it conformal 
weight} $h$. 
In an irreducible representation of the Virasoro algebra 
the {\it central extension} operator $C$ has a fixed value $c\in \bbbc$
since it commutes with the
whole Virasoro algebra. Therefore it is common practice
to omit $c$ in the highest weight
vector $\ket{h,c}$ and consider it as a fixed constant.
We construct the freely generated module $V_{h,c}$ on a highest
weight vector $\ket{h,c}$ which is called the {\it Verma module} of 
$\ket{h,c}$. 
A basis for $V_{h,c}$ is given by
\bea
B_{h,c} &=& \Bigl\{ L_{-n_{i}} L_{-n_{i-1}} \ldots L_{-n_{2}} 
L_{-n_{1}} \ket{h,c} : n_{i} \geq \ldots \geq n_{1} , \bk
n_{j} \in \bbbn, \bk i\in\bbbno \Bigr\} \pkt \label{eq:virbas} 
\eea
The action of the Virasoro algebra on $V_{h,c}$ is simply given by its 
commutation relations and the action on the highest weight state 
$\ket{h,c}$. 
By defining the {\it triangular decomposition}
of the Virasoro algebra, which we
shall call the algebra $\scn{0}$, we can write $V_{h,c}$ using tensor 
product notation:
\bea
& {\scn{0}} \;=\; {\scn{0}^{-}} \oplus {{\cal H}_{0}} \oplus 
{\scn{0}^{+}} 
\com \\
& \scn{0}^{\pm} \;=\; \spn{L_{\pm n}: n\in \bbbn} \com \sk 
{\cal H}_{0} \; = \; \spn{L_{0},C} \com \nn \\
& V_{h,c}\;=\;
U(\scn{0})
{\displaystyle\otimes_{{\cal H}_{0}\oplus\scn{0}^{+}}} \ket{h,c} \com
\eea
where $U(\scn{0})$ denotes the {\it universal enveloping algebra} of
the Virasoro algebra $\scn{0}$. The 
Verma module $V_{h,c}$ decomposes into a direct sum of
$L_{0}$ grade spaces $V_{h,c}^{n}$ with the basis
\bea
B^{n}_{h,c} \! &\!=\!& \! \Bigl\{ L_{-n_{i}} L_{-n_{i-1}} \ldots L_{-n_{2}} 
L_{-n_{1}} \ket{h,c} : n_{i} \geq \dots \geq n_{1} , \bk n_{i} + \ldots 
+n_{1} = n , \bk 
n_{j} \in \bbbn \Bigr\} \, , \label{eq:bnhc}
\eea
where $B^{0}_{h,c}$ is defined to be $\{ \ket{h,c} \}$.
If $R_{h,c}$ is an irreducible highest weight representation 
of the Virasoro algebra with highest weight $h$ and 
fixed value $c$ for $C$, then there exists a homomorphism $\phi_{h,c}$ from 
$V_{h,c}$ onto $R_{h,c}$. If $V_{h,c}$ is reducible, then the kernel 
$K_{h,c}$ of $\phi_{h,c}$ is non-trivial. 
It can be shown that in this case $K_{h,c}$ can also
be decomposed in $L_{0}$ grade spaces $K^{n}_{h,c}$. If a vector lies in 
$K_{h,c}$ it is obvious that all its {\it descendant vectors}, 
obtained by acting with Virasoro operators of negative index on the vector
and taking linear combinations, 
also lie in $K_{h,c}$. Hence, if $K_{h,c}^{0}$ is non-trivial we find that 
the homomorphism $\phi_{h,c}$ is trivial and we obtain the trivial 
representation. If $R_{h,c}$ is not the trivial representation we have 
$K^{0}_{h,c}=\{ 0\}$. Thus there exists a smallest index $j \in \bbbn$ 
such that $K^{j}_{h,c}$ is non-trivial. If we take $\psi \in K^{j}_{h,c}$ 
and 
$L_{m}$ with positive index, then $\phi_{h,c}(L_{m}\psi) = L_{m} 
\phi_{h,c}(\psi) =0$ 
and hence due to the minimality of $j$ we find $L_{m} \psi = 
0$. The vector $\psi$ is not proportional to the highest weight vector 
$\ket{h,c}$ but satisfies highest weight vector conditions with highest 
weight $h+j$. We call such a vector a {\it singular vector}
in $V_{h,c}$ at level $j$:
a vector $\psi_{n} \in V_{h,c}$ is called singular vector at level
$n$ if
\bea
L_{0} \psi_{n} = (h+n) \psi_{n} \com &&
L_{m} \psi_{n} = 0 \com \sk \forall m \in \bbbn \pkt
\eea
For the Virasoro algebra it can be shown\cite{ff1}
that any vector in the kernel 
$K_{h,c}$ is either a descendant of a singular vector or is singular 
itself. For all algebras, 
a highest weight representation is irreducible if and 
only if there are no singular vectors in the representation. This is 
fundamental to understand the 
significance of the singular vectors. Furthermore, as proven by 
Feigin and Fuchs\cite{ff1}, if 
we know in the Virasoro case 
the singular vectors in $V_{h,c}$ we can construct the irreducible 
representation $R_{h,c}$ by acting on $V_{h,c}$ with a 
homomorphism whose 
kernel consists of the sum of the submodules spanned by 
the singular vectors and their 
descendants\footnote{For other algebras,
after acting with the homomorphism which puts all
singular vectors and their descendants equal to zero in the Verma module, 
new singular vectors may appear which were initially not singular in the
Verma module. Such vectors are the so-called {\it subsingular vectors}
of the Verma module.}. 
The structure of the highest weight representations 
of \eq{\ref{eq:vir}} are by now very well understood thanks to
the combined effort of several 
authors\cite{bauer1,bauer2,bsa1,bsa2,ff1,friedan2,fuchs,gko,kac,adrian1}.

So far, we have been looking at a theory describing a
conformally invariant physical model at algebraic level. 
The underlying quantum field theory 
contains the quantum fields $\Phi_{h}(z)$ which
generate the energy eigenstates $\ket{h,c}$
from the vacuum: $\ket{h,c}=\Phi_{h}(z)\ket{0}_{c}$.
Here we fixed again the central extension term: $C\ket{0}_{c}
=c\ket{0}_{c}$. These
fields are called the {\it primary fields}. 
The field generating the conformal transformations and hence having the
Virasoro generators as modes, is the stress-energy tensor $T(z)$:
\bea
T(z) &=& \sum_{n\in\bbbz} (z-w)^{-n-2} L_{n}(w) \pkt
\eea
As we explain later in the context of superconformal field theory,
performing two conformal 
transformations and
using contour integration methods allow us to compute the commutator of
algebra generators and primary fields:
\bea
[L_{m}(w),\Phi_{h}(z)] &=& \Bigl[ h(m+1)(z-w)^{m}+(z-w)^{m+1}\frac{\dif}{
\dif z}\Bigr] \Phi_{h}(z) \pkt \label{eq:vircomphi}
\eea
We can always shift any point $z$ to the origin
since the group of translations is contained in the group of conformal 
transformations. 
Therefore we denote $L_{m}(0)$ simply by $L_{m}$ and we take the vector
$\ket{h,c}$ as generated at the origin: $\ket{h,c}=\Phi_{h}(0)\ket{0}_{c}$.
We can give a complete set of fields of the conformally invariant
theory by acting with
the modes of $T(z)$ on primary fields $\Phi_{h}(w)$
to obtain the {\it descendant fields} of $\Phi_{h}(w)$. 

Singular vectors vanish in
the physical theory. Therefore, correlators with
singular vector operators inserted have to vanish.
Using \eq{\ref{eq:vircomphi}} one can thus obtain differential equations
for the correlators of the theory
by inserting singular operators. Hence, singular vectors together with 
\eq{\ref{eq:vircomphi}} describe the
dynamics of the physical model. For this reason we need to know
not only singular vectors and thus irreducible representations, but also
the action of the algebra generators on the primary field as this describes
the dynamics. In this paper we will focus on the definition of primary fields
in superspace using a superfield formalism. These are exactly the theories 
which are known as the {\it Neveu-Schwarz} theories. 

In a physical model for elementary particles which has a Lie algebra 
as symmetry generators, the statistics of the particles are left unchanged
under the action of this algebra. 
However, there is a common belief that a {\it theory of everything} should 
have a symmetry, transforming particles of different statistics into
one another and hence providing a geometrical framework in which
{\it fermions} and {\it bosons} receive a common treatment.
Such a symmetry can be realised using a symmetry algebra which is
$\bbbz_{2}$-graded in the sense that some of their elements satisfy
anticommutation relations rather than commutation relations
and the underlying geometry can be provided by supermanifolds. These
$\bbbz_{2}$-graded algebras form
{\it Lie superalgebras}. Motivated not only by 
string theory but also by two-dimensional statistical critical
phenomena, the Lie superalgebra extensions of the Virasoro algebra
became very attractive,
as first suggested by Ademollo {\it et al.}\cite{ademello}. 
At the same time Kac\cite{kacLieSAlg} independently constructed
several series of simple infinite-dimensional Lie superalgebras,
among them superextensions of the Virasoro algebra.
Since then,
many applications for {\it superconformal
field theory} were found, not only of theoretical interest. 
The {\it tricritical Ising model}, 
which can be realised experimentally\cite{ferreira},
was identified by
Friedan, Qiu, and Shenker\cite{friedan} 
as a $N=1$ superconformal model (we will reveal later 
the significance of the parameter $N$ in that context). 
Moreover, the $N=2$ superconformal models 
find applications in
critical phenomena since under certain circumstances $O(2)$ Gaussian
models are $N=2$ superconformally invariant\cite{waterson}.
There has recently been great interest in superconformal 
field theories because of their applications in superstring
theory.  
The $N=2$ superstring seems to be particularly
interesting because of its connexion to quantum
gravity\cite{nicolai,nishino1,ooguri} 
and two-dimensional black holes. Furthermore it
has been conjectured that the $N=2$ string should give us
insight into integrable systems\cite{atiyah}.
Just recently, Kac has proven a complete classification
of superconformal algebras\cite{kacCMP2}.

After setting up the necessary supergeometric framework in \secn{2},
we define in \secn{3} the notion of superconformal transformations.
In \secn{4} and \secn{5} we construct the foundation of
superintegration and super Taylor expansions. This enables us to define
in \secn{6} superconformal field theories and to derive
the well-known examples of $N=1$ and $N=2$ superconformal field theories
in \secn{7} and \secn{8}.
The Hilbert space of states of a conformal field theory
is created by the action of superconformal primary fields on the vacuum
state and furthermore the action of the whole superconformal algebra on 
these highest weight vectors. Therefore the action of the algebra
generators on the superconformal primary fields is of particular interest.
We investigate the global transformation properties of uncharged
superconformal primary fields in \secn{9}.
Like a conformal field theory, a superconformal field theory consists
of two chiral sectors having equivalent representation theories.
For this reason we will restrict our definition to one 
chiral sector only: the holomorphic part. We therefore
leave the antiholomorphic coordinates always unchanged and 
omit them in the notation.

\section{Supergeometry}

As pointed out earlier, the ideas of having symmetry algebras
which transform particles of different statistics into one another
requires the extension of Lie algebras by anticommuting objects.
This can be done by extending a Lie algebra to a $\bbbz_{2}$-graded algebra
which defines the notion of a Lie superalgebra. 
The theory of Lie superalgebras is well established in the 
mathematical literature and we certainly do not want to
rederive this here. The interested reader
will find a vast amount of literature on
this topic among which we want to point out the paper
by Kac\cite{kacLieSAlg} and the book by Scheunert\cite{scheunert}.
For our purpose we shall give a more simplified definition of a Lie
superalgebra starting
already from an associative algebra:

\bdf Lie superalgebra\\
Consider a $\bbbz_{2}$-graded associative algebra $A=A_{0} \oplus A_{1}$
with $a_{p} b_{q} \in A_{p+q \: (\mod \, 2)}$ for $a_{p} \in A_{p}$, $b_{q} \in
A_{q}$ and $p,q \in \{ 0,1 \}$.
We define the bilinear 
supercommutator by defining it for $a_{p} \in A_{p}$ and $b_{q} \in
A_{q}$, $p,q \in \{ 0,1 \}$:
\bea
[a_{p},b_{q}]_{S}&=&a_{p}b_{q}-(-1)^{pq}b_{q}a_{p} \pkt \nn 
\eea
$A$ is called Lie superalgebra. The elements of $A_{0}$ are qualified as
even and the ones of $A_{1}$ as odd. 
\end{definition}

We have now defined what in a supertheory will play the r\^{o}le of the 
symmetry algebra. However, in order to define a quantum field theory, we need
to define the underlying manifold. 
The concept of supermanifolds, extensions of differential manifolds,
is well understood. 
Among other references we shall point out the book by Manin\cite{manin}.
However, we want to achieve the extension
of Riemann surfaces, the underlying manifolds of conformal field theories.
A subclass of these
so-called superconformal manifolds, also known as super-Riemann surfaces, 
were first studied by Friedan\cite{friedan3}. This was later generalised by 
Cohn\cite{cohn}.
The definition we give here claims by 
no means to be exhaustive but should rather be understood as
an incentive.

In order to 
extend an ordinary quantum field theory to a super quantum field 
theory with underlying {\it supermanifold} one
would construct a fibre bundle of anticommutative rings over the manifold
of the model. 
As far as coordinates are concerned we obtain 
the ones of the 
manifold plus anticommuting Grassmann variables arising due to the attached
anticommutative rings.

\bdf Anticommutative ring\\
An algebra $R$ over the complex numbers $\bbbc$
is called anticommutative ring if it is $\bbbz_{2}$-graded
$R=R_{0} \oplus R_{1}$ such that $a_{p}b_{q}\in R_{p+q \: (\mod \, 2)}$
for $a_{p}\in R_{p}$ and $b_{q}\in R_{q}$ where $p,q \in \{ 0,1 \}$.
Moreover the bilinear 
supercommutator is trivial:
\bea
[a_{p},b_{q}]_{S} &=\: a_{p}b_{q} -(-1)^{pq} b_{q} a_{p} \: =0 \com 
\nn 
\eea
with $a_{p} \in R_{p}$ and $b_{q} \in
R_{q}$ for $p,q \in \{ 0,1 \}$. Furthermore we require that there exists
a generating\footnote{Including the zero-power product,
i.e. the identity.} set $R_{\theta} \subset R_{1}$. 
\edf
The elements of $R_{\theta}$ shall be
called {\it Grassmann} variables. They satisfy
\bea
\theta_{1} \theta_{2} &=& -\theta_{2}\theta_{1} \com \sk \theta_{1},\theta_{2}
\in R_{\theta} \com
\eea
and since they generate $R$, we can write $R$ as the ring of polynomials over
$\bbbc$ generated by $R_{\theta}$: $R=\bbbc[R_{\theta}]$.

The number of anticommutative rings we tensor together in the fibres has
to be the same for the whole manifold.
It is called the
{\it classification parameter} $N$ of the supermanifold. 
The theories we aim to construct
are based on the manifolds of conformal field theories, 
more precisely on
Riemann surfaces having the complex coordinate $z$.
To construct the super extension we take $N$ anticommutative rings
in its fibres. We obtain the set of coordinates $(z,\theta_{1},
\ldots ,\theta_{N})$ and we then extend the complex differential structure 
by anticommuting derivatives $\frac{\partial}{\partial \theta_{i}}$:
\bea
\pder{}{\theta_{i}}\pder{}{\theta_{j}} &=&
-\pder{}{\theta_{j}}\pder{}{\theta_{i}} \com \bk\bk
\pder{}{\theta_{i}}\theta_{j}=\delta_{i,j} -
\theta_{j} \pder{}{\theta_{i}} \pkt
\eea 
Finally, we define the 
{\it superderivatives} $D_{i}=\frac{\partial}{\partial \theta_{i}} +
\theta_{i} \frac{\partial}{\partial z}$ for 
$i=1,\ldots , N$. The superderivatives are the square roots of the 
complex derivative $\partial_{z}=D_{i}^{2}$, 
and therefore they describe exactly the fermionic structure we
expected. Performing a coordinate transformation 
from $(z,\theta_{1}, \ldots, \theta_{N})$ to
$(\bar{z},\bar{\theta}_{1}, \ldots, \bar{\theta}_{N})$ the
superderivatives transform as follows\footnote{The usual summation 
convention applies.}:
\bea
D_{i}&=&\underbrace{(D_{i} \bar{\theta}_{j}) \bar{D}_{j}}_{
\rm homogeneous\;\; part} + 
\bk \underbrace{(D_{i}
\bar{z} - \bar{\theta}_{j} D_{i} \bar{\theta}_{j})
\partial_{\bar{z}}}_{\rm inhomogeneous\;\; part} \pkt 
\label{eq:supdertrans}
\eea
In a conformal field theory the derivatives transform covariantly 
obtaining a prefactor only: $\frac{\partial}{\partial z} =
\frac{\partial \bar{z}}{\partial z} \frac{\partial}{\partial \bar{z}}$. 
We apply the analogue statement for the square roots $D_{i}$ of
$\partial_{z}$ to define a superconformal field theory.
Among all supertransformations we pick those which 
transform the superderivatives by a scaling factor only
and which are therefore conformal in both the even and the odd variables. 
Hence, we
require the inhomogeneous part in \eq{\ref{eq:supdertrans}} to vanish:
\bdf Superconformal transformations\\
A transformation from $(z,\theta_{1}, \ldots, \theta_{N})$ to
$(\bar{z},\bar{\theta}_{1}, \ldots, \bar{\theta}_{N})$ 
is called superconformal if
\bea
D_{i}&=&(D_{i}\bar{\theta}_{j})\bar{D}_{j} \com \bk
 1 \leq i \leq N  \pkt \label{eq:supconf} 
\eea 
\edf
We conclude this section by defining the underlying manifold
of a superconformal field theory:
\bdf Superconformal manifold\\
A superconformal manifold $\pmb{${\cal S}$}_{N}$ of classification
parameter $N$ is a fibre bundle of $N$ anticommutative rings over a
one-dimensional complex manifold where the transition functions are
superconformal transformations. The coordinates shall be called
superpoints $Z=(z,\theta_{1},\ldots ,\theta_{N})$.
\edf
The space of functions $\pmb{${\cal F}$}_{N}$ defined on a superconformal
manifold consists of functions 
$f(Z)=f^{0}(z) + \theta_{i} f_{i}^{1}(z) + \ldots +
\theta_{1} \theta_{2} \ldots \theta_{N} f^{N}(z)$ where the functions 
$f^{0}(z)$, $\ldots$, $f^{N}(z)$ are complex functions evolving
according to superconformal transformations.


\section{Superconformal transformations} \label{sec:supconf}

Following the original approach of Kac\cite{kacLieSAlg} 
we define the differential
form $\kappa = \dif z + \theta_{i} \dif \theta_{i}$. 
Superconformal
transformations are the only transformations under which 
$\kappa$ will simply be scaled. 
We find the following equivalences:
\bth
\bea
& D_{i}=(D_{i}\bar{\theta}_{j})\bar{D}_{j} \com  \forall i \bk
\Leftrightarrow \bk D_{j}  \bar{z} = \bar{\theta}_{i} D_{j}
\bar{\theta}_{i} \bk \com \forall j \bk \Leftrightarrow \bk
\bar{\kappa} = p \kappa \com \label{eq:formequiv}
\eea
where the prefactor $p(z,\theta_1,\ldots,\theta_N)$ is given by 
\bea
p&=&\pder{\bar{z}}{z} + \bar{\theta}_{i}
\pder{\bar{\theta}_{i}}{z} \pkt \nn 
\eea
\eth
\bprf
The transformation of $\kappa$ is
\bea
\bar{\kappa}&=& \left( \pder{\bar{z}}{z}+\bar{\theta}_{i}
\pder{\bar{\theta}_{i}}{z}\right) \dif z 
+ \left( - \pder{\bar{z}}{\theta_{j}} +
\bar{\theta}_{i}\pder{\bar{\theta}_{i}}{\theta_{j}} 
\right) \dif \theta_{j} \pkt \label{eq:kaptrans}
\eea
$\bar{\kappa}=p\kappa$ implies $\left( \pder{\bar{z}}{z}+\bar{\theta}_{i}
\pder{\bar{\theta}_{i}}{z}\right)\theta_{j}=$
$\left( - \pder{\bar{z}}{\theta_{j}} +
\bar{\theta}_{i}\pder{\bar{\theta}_{i}}{\theta_{j}} 
\right)$ which is consequently equivalent to $D_{j}\bar{z}=
\bar{\theta}_{i}D_{j}\bar{\theta}_{i}$. Thus the scaling factor $p$
can be found in \eq{\ref{eq:kaptrans}}.
\eprf

Hence, finding the generators of superconformal transformations
is equivalent to finding
the superderivatives acting on the space of
differential forms ${\cal D}=\bbbc [z,z^{-1}] \otimes_{\bbbc}
 \bbbc [ d \,
z] \otimes_{\bbbc} \bbbc [\theta_{1},\ldots,\theta_{N}]
\otimes_{\bbbc} \bbbc [d\, \theta_{1},\ldots ,d\, \theta_{N} ]$ and
leaving $\kappa \in {\cal D}$ invariant up to a scalar
multiple, i.e. $\bar{\kappa}=p
\kappa$ for some $p$ depending on the coordinates.
For our further considerations we present the result of Kac\cite{kacLieSAlg} 
in the form recently given by Bremner\cite{bremner}.  One takes
elements $i_{1}, \ldots , i_{I} \in \{ 1, \ldots , N\}$ which form
the sequence $S=(i_{1}, \ldots , i_{I})$. In addition one defines the 
complement of $S$ as a set $\bar{S}=\{1,\ldots,N\} \without S$ and finally
constructs operators labeled by a sequence $S$
and an index $a$ which is taken from $\bbbz$ if the 
number of elements in $S$ is even or otherwise $a$ is taken from $\bbbzh$:
\bea
X_{a}(i_{1},\ldots,i_{I})&=& (1-\frac{I}{2}) z^{a-\frac{I}{2}+1} 
\theta_{i_{1}} \ldots \theta_{i_{I}} \partial_{z}
+\half \sum_{p=1}^{I} (-1)^{p+I} z^{a-\frac{I}{2}+1} 
\theta_{i_{1}} \ldots \check{\theta}_{i_{p}} \ldots \theta_{i_{I}} 
\partial_{\theta_{i_{p}}} \nn \\
&& +\half (a-\frac{I}{2}+1) \sum_{k\in\bar{S}} z^{a-\frac{I}{2}}
\theta_{i_{1}} \ldots \theta_{i_{I}} \theta_{k} \partial_{\theta_{k}}
\com \label{eq:sgen}
\eea
where $\check{\theta}_{i_{p}}$ signifies that $\theta_{i_{p}}$ is omitted
in the product.
$X_{a}(i_{1},\ldots,i_{I})$ is defined to be even if $a\in\bbbz$,
otherwise it is qualified as odd.
A basis for the space of operators 
leaving $\kappa$ invariant up to a scalar multiple is given by the set of
$X_{a}(i_{1}, \ldots , i_{I})$ with $1\leq i_{1}< \ldots
<i_{I}\leq N$. 
The operators \eqoth{\ref{eq:sgen}} satisfy the supercommutation relations
\bea
[X_{a}(i_{1},\ldots,i_{I}),X_{b}(j_{1},\ldots,j_{J})]_{S} \!\!&=&\!\!
\sum_{p=1}^{I} \sum_{q=1}^{J} 
\frac{(-1)^{I+p}\delta_{i_{p},j_{q}}}{2}
X_{a+b}(i_{1},\ldots,\check{i}_{p},\ldots,i_{I},
j_{1},\ldots,\check{j}_{q},\ldots,j_{J}) \nn \\
&& + [(1-\frac{I}{2})b-(1-\frac{J}{2})a] X_{a+b}(i_{1},\ldots,
i_{I},j_{1},\ldots,j_{J}) \com \label{eq:scom}
\eea 
which are closed in the set of basis elements by 
reordering\footnote{Note
that $X_{a}(i_{1},\ldots,i_{I})$ is trivial if $(i_{1},\ldots,i_{I})$
contains the same element twice or otherwise
it is proportional to the basis element 
$X_{a}(i^{\prime}_{1},\ldots,i^{\prime}_{I})$ with 
$(i^{\prime}_{1},\ldots,i^{\prime}_{I})$ being 
$(i_{1},\ldots,i_{I})$ reordered appropriately.}
the union of the
sequences $(i_{1},\ldots,i_{I})$ and $(j_{1},\ldots,j_{J})$.
Moreover for the transformation
of $\kappa$ one can find the scaling factor:
\bea
\bar{\kappa} \:=\: X_{a}(i_{1},\ldots,i_{I})\kappa &=&
(a-\frac{I}{2}+1)z^{a-\frac{I}{2}} \theta_{i_{1}} \ldots
\theta_{i_{I}} \kappa \pkt
\eea

The result by Kac gives the symmetry generators of a {\it classical
superconformally} invariant field theory. In particular for $N=0$ we
obtain a conformally invariant classical model having the 
de Witt algebra as symmetry algebra. It contains the operators 
$X_{a}=z^{a}\partial_{z}$ where $a\in\bbbz$, satisfying the commutation 
relations
\bea
[X_{a},X_{b}]&=&(b-a) X_{a+b} \pkt
\eea
It is a feature of infinite dimensional Lie algebras to
allow in the quantised theory {\it central terms} which extend
the classical symmetry algebra but do not change
the infinitesimal transformations of tensors. In the case of $N=0$ 
the allowed central term leads us to the Virasoro algebra.
Later we shall find that for the 
superconformal theories we have
as well at most one central term which, however, disappears completely
if $N$ is at least equal to 4.


\section{Superconformal integration}

\subsection{Superdifferentials}
 
We have found the superconformal analogues $D_{i}$ of the conformal
derivatives $\der{}{z}$. In order to find the analogue of the radial
quantisation procedure used for two-dimensional conformal field theories,
we have to develop the corresponding integration and 
contour integration techniques. 
In order to do so we need to define differentials $\dif Z_{j}$ as duals
of $D_{i}$.
\bdf Superdifferentials\\
We define the differentials $\dif Z_{i}$, $i=1,  \ldots ,N$, as the
dual elements of the superderivatives $D_{i}$:
\bea
D_{i} \, \dif Z_{j} = \delta_{i,j} \com \bk\bk i,j \in \{ 1, \ldots , N \} 
\pkt 
\eea
\edf

If we are given a superconformal transformation\footnote{Due
to \eqs{\ref{eq:formequiv}} the derivatives of $\bar{z}$ are 
determined by $\bar{\theta}_{1}, \ldots, \bar{\theta}_{N}$ and hence
contain no linearly independent information.}
$\bar{\theta}_{1}(
z,\theta_{1},\ldots ,\theta_{N}),\ldots,\bar{\theta}_{N}(
z,\theta_{1},\ldots ,\theta_{N})$ then the matrix of
superderivatives will contain all the information of directional
derivatives and hence be crucial to define integration.
\bdf Super Jacobi matrix\\
We define the matrix of superderivatives
\bea
D\bar{\theta} & = &
\left(
\begin{array}{ccc} 
D_{1}\bar{\theta}_{1} & \ldots & \ D_{N} \bar{\theta}_{1} \\
\vdots & & \vdots \\
D_{1}\bar{\theta}_{N} & \ldots & \ D_{N} \bar{\theta}_{N}
\end{array} \right) \com \label{eq:superjakob}
\eea
which we can write in index notation as
$(D\bar{\theta})_{i,j}=D_{j} \bar{\theta}_{i}$.
\edf
As the first fruitful result of these definitions we can check easily that
the usual chain rule holds:
\bea
D \: \bar{\!\bar{\theta}} &=& ( \bar{D} 
\: \bar{\!\bar{\theta}}) ( D \bar{\theta}) \pkt \label{eq:chainrule}
\eea
We thus obtain the transformation rule for
the vector of differentials $\dif Z = (\dif Z_{1} ,\ldots , 
\dif Z_{N})^{T}$ under superconformal transformations:
\bea
\dif \bar{Z} &=& D \bar{\theta} \dif Z \pkt \label{eq:diftrans}
\eea

\subsection{Riemann superintegrals}
We say $F(Z)$ is a {\it $Z_{i}$-integral} of $f(Z)$ if
$D_{i} F(Z) = f(Z)$. In symbols we write
\bea
F(Z) &=& \int \dif Z_{i} f(Z) \pkt
\eea
Obviously, $D_{i} F(Z) =0$  
with $F(Z)=F^{0}(z) + \theta_{j} F_{j}^{1}(z) + \ldots +
\theta_{1} \theta_{2} \ldots \theta_{N} F^{N}(z)$ and 
fixed $i$ implies
that the only possible non-trivial components of $F(Z)$ 
are those which are not a coefficient of the coordinate $\theta_{i}$.
Furthermore, their derivative $\pder{}{z}$ has to vanish and 
they are therefore constant. Thus, $F(Z)$ is constant in 
$\theta_{i}$ direction. 
This implies automatically that due to the linearity of the
differential operator $D_{i}$ two $Z_i$-integrals of $f(Z)$
differ at most by a factor which is
constant if we fix $\theta_{j}$ for $j=1,\ldots, 
N$, $j\neq i$. 
Defining integration
over the superdifferentials $\dif Z_{i}$ using a generalisation
of the fundamental theorem of calculus is therefore justified.
\bdf Riemann superintegrals\\
We define the Riemann superintegral of a function f(Z) as
\bea
\int_{Z_{1}}^{Z_{2}} \dif Z_{i} \, f(Z) &=& F(Z_{2}) -F(Z_{1}) \com
\eea
where $F(Z)$ is a $Z_i$-integral of $f(Z)$, $i=1,\ldots , N$,
and the superpoints $Z_{1}$ and $Z_{2}$ coincide with the possible exceptions
of their $z$ and $\theta_{i}$ coordinates.
\edf
In the view of further applications we 
define for two superpoints\footnote{One
should not confuse the component index with the label index of superpoints,
superderivatives and superdifferentials.} $Z_{i}=
(z_{i},\theta_{i,1},\ldots , \theta_{i,N})$, $i\in \{1,2\} $,
the differences $Z_{12}=z_{1}-z_{2}-\theta_{1,j} 
\theta_{2,j}$ and $\theta_{12,j}=\theta_{1,j} -
\theta_{2,j}$. The importance of these 
superdifferences\footnote{$(i)$ denotes no summation convention applies to 
the index $i$.} lies in
$D_{2,i} Z_{12}^{n} = n \theta_{12,i} Z_{12}^{n-1}$ and
$D_{2,(i)} \theta_{12,(i)} Z_{12}^{n} = - Z_{12}^{n}$. 
This means that they are the successive $Z_i$-integrals of 1:
\bea 
\int_{Z_{1}}^{Z_{2}} \dif Z_{3,i} Z_{13}^{n} &=& - 
\theta_{12,i} Z_{12}^{n} \com \label{eq:unit1} \\
\int_{Z_{1}}^{Z_{2}} \dif Z_{3,(i)} \theta_{13,(i)} 
Z_{13}^{n} &=& \frac{1}{n+1} Z_{12}^{n+1} \pkt \label{eq:unit2}
\eea
Moreover, we now define integrals over a volume of the superconformal
space:
\bdf Superconformal integral over a volume\\
We take $f(Z)\in \pmb{${\cal F}$}_{N}$ 
and let $V$ be a volume in the superconformal space
$\pmb{${\cal S}$}_{N}$. We then define the superconformal
integral over volume $V$:
\bea
\int_{V} \dif Z f(Z) &=& \int_{V} \dif Z_{1} \ldots \dif Z_{N} f(Z)
\label{eq:volint}
\pkt  
\eea
\edf
The chain rule \eqoth{\ref{eq:diftrans}} leads us to the substitution rule
for integrals of the form \eqoth{\ref{eq:volint}}.
We take a scalar function $f(Z)$ of the superpoint $Z$ and 
perform a superconformal transformation\footnote{Note 
that we do not consider any superdeterminants. $D
\bar{\theta}$ is merely an even object.} $Z \mapsto \bar{Z}$:
\bea
\int_{V} \dif Z f(Z) &=&
\int_{\bar V} \dif \bar{Z} f(\bar{Z})  \det  D \bar{\theta} \pkt
\eea

\subsection{Supercontour integrals}

Finally we want to define contour integrals on the superconformal 
manifold $\pmb{${\cal S}$}_{N}$.
Whilst for Riemann superintegrals we aimed to find a function which has a
given function as derivative and satisfies certain boundary conditions,
for contour integrals we are more interested in defining an
extension which is translation invariant and linear just like
ordinary contour integrals.
These two properties fix the contour integrals already up to a
scalar factor. We follow the standard approach to define contour integration
over Grassmann variables.
\bdf
We define contour integrals over $\theta_{j}$ as
\bea
\oint_{{\cal C}_{0}} d \theta_{i} \bk \theta_{j} &=& \delta_{i,j} \com \\
\oint_{{\cal C}_{0}} d \theta_{i} \bk 1           &=& 0       \com \nn 
\eea
where ${\cal C}_{0}$ is a supercontour about the origin.
\edf
These simple integration rules have the effect that for a
function $f(Z)\in\pmb{${\cal F}$}_{N}$ the only contributing
term towards the contour integral is $f^{N}(z)$, due to: 
\bea
\oint_{{\cal C}_{0}} \dif \theta_{1} \ldots \dif \theta_{N} \bk
\theta_{1} \ldots \theta_{N} &=&
(-1)^{\frac{N(N-1)}{2}} \com \label{eq:grasint}
\eea
and the integral vanishes whenever some of the $\theta_{i}$ are
missing in the product $\theta_{1} \ldots \theta_{N}$. 
This leads to the definition:
\bdf Supercontour integrals \label{def:scontint}\\
For the function
$f(Z)=f^{0}(z) + \theta_{i} f^{1}_{i}(z) + \ldots +  
\theta_{1} \ldots
\theta_{N} f^{N}(z)\in\pmb{${\cal F}$}_{N}$
we define the integral along the supercontour
${\cal C}$ as
\bea
\oint_{\cal C} \dif Z f(Z) &=& \oint_{\cal C} \dif z \dif \theta_{1} 
\ldots \dif \theta_{N}
f(Z)
\nn \\
&=& \epsilon^{N} \oint_{{\cal C}^{\prime}} \dif z f^{N}(z) \com 
\eea
where $\epsilon^{N} = (-1)^{\frac{N(N-1)}{2}}$, and ${\cal C}^{\prime}$
is the projection of the supercontour $\cal C$ into the underlying
Riemann surface.
\edf


\section{Super Taylor expansion}

We already know from the definition of a function
on a superconformal space how it can be expanded in a power
series about the origin. In this section we derive
an expansion about non-trivial superpoints. We thus obtain 
an expansion in terms of the superdifferences $Z_{12}$ and 
$\theta_{12,j}$.

\bth Super Taylor expansion\\
The super Taylor expansion of $f(Z)\in \pmb{${\cal F}$}_{N}$ is given by
\bea
f(Z_{1}) &=& \sum_{n=0}^{\infty} \frac{1}{n!} Z_{12}^{n}
\partial_{z_{2}}^{n} \prod_{j=1}^{N} (1+\theta_{12,j} D_{2,j})
f(Z_{2}) \label{eq:taylor} \pkt
\eea 
\eth
\bprf
We first consider the case\footnote{Note
that for odd functions $f$ we can integrate
by parts by altering the signs:
$\int \dif Z \; [Df(Z)] g(Z) = 
f(Z) g(Z) + \int \dif Z \; f(Z) [D g(Z)]$. If $f(Z)$ is even we can 
integrate by parts as usual.}
$N=1$: $Z_{i}=(z_{i},\theta_{i})$, 
$i \in \{ 1,2\} $.
\bea
f(Z_{2}) &=& f(Z_{1})+ \int_{Z_{1}}^{Z_{2}} \dif Z Df(Z) \nn \\
&=& f(Z_{1}) + \left. ( \int_{Z_{1}}^{Z} dZ^{\prime} ) Df(Z)
\right|_{Z_{1}}^{Z_{2}} + \int_{Z_{1}}^{Z_{2}} \dif Z ( \int_{Z_{1}}^{Z}
dZ^{\prime} ) D^{2} f(Z) \nn \\
&=&f(Z_{1}) + \left. (\theta - \theta_{1}) D f(Z)
\right|_{Z_{1}}^{Z_{2}} + \int_{Z_{1}}^{Z_{2}} dZ (\theta -
\theta_{1}) D^{2} f(Z) \nn\\
&=& f(Z_{1}) - \theta_{12} D_{2} f(Z_{2}) + \left. [\int_{Z_{1}}^{Z}
\dif Z^{\prime} (\theta^{\prime} -\theta_{1})] D^{2}f(Z)
\right|_{Z_{1}}^{Z_{2}} \nn \\ && -
\int_{Z_{1}}^{Z_{2}} \dif Z (z-z_{1}-\theta \theta_{1}) D^{3} f(Z) \nn \\
& \vdots & \nn\\
&=& f(Z_{1}) - \theta_{12} D_{2} f(Z_{2})-Z_{12} D^{2}_{2}
f(Z_{2}) -\frac{1}{2}
\theta_{12} Z_{12} D^{3}_{2}
 f(Z_{2}) \nn \\ && - \frac{1}{2} Z_{12}^{2} D^{4}_{2}
f(Z_{2}) - \frac{1}{3!} Z_{12}^{2} \theta_{12} D^{5}_{2} f(Z_{2}) - \ldots
\pkt \nn 
\eea
Using \eqs{\ref{eq:unit1}} and \eqoth{\ref{eq:unit2}} we can easily prove
by induction:
\bea
f(z_{1},\theta_{1}) & = & \sum_{n=0}^{\infty} \frac{1}{n!} Z_{12}^{n}
(1+ \theta_{12} D_{2}) D_{2}^{2n} f(Z_{2}) \nn \\
&=& \sum_{n=0}^{\infty} \frac{1}{n!} Z_{12}^{n} \partial_{z_{2}}^{n}
(1+ \theta_{12} D_{2}) f(Z_{2}) \pkt \label{eq:N1tayp}
\eea
For the general case we define the sequence of superpoints:
\bea
Z_{1}^{0}&=&(z_{1},\theta_{1,1},\theta_{1,2},\ldots,\theta_{1,N}) \com \nn\\
Z_{1}^{1}&=&(z_{2},\theta_{2,1},\theta_{1,2},\ldots,\theta_{1,N}) \com \nn\\
&\vdots \nn\\
Z_{1}^{i}&=&(z_{2},\theta_{2,1},\ldots,\theta_{2,i},\theta_{1,i+1}, 
\ldots,\theta_{1,N}) \com \nn\\
&\vdots \nn \\
Z_{1}^{N}&=&(z_{2},\theta_{2,1},\theta_{2,2},\ldots,\theta_{2,N}) \pkt 
\nn
\eea
We apply \eq{\ref{eq:N1tayp}} which was found for $N=1$:
\bea
f(Z_{1}^{0})&=& \sum_{n=0}^{\infty} \frac{1}{n!} (Z_{12}^{1})^{n}
\partial_{z_{2}}^{n} (1+ \theta_{12,1} D_{2,1}) f(Z_{1}^{1}) \nn\\
&=& \sum_{n=0}^{\infty} \frac{1}{n!} (Z_{12}^{1})^{n}
\partial_{z_{2}}^{n} (1+\theta_{12,1} D_{2,1}) \sum_{m=0}^{\infty}
\frac{1}{m!} (Z_{12}^{2})^{m} \partial_{z_{2}}^{m} (1+ \theta_{12,2}
D_{2,2}) f(Z_{1}^{2}) \com \nn 
\eea
where $Z_{12}^{1}=z_{1}-z_{2}-\theta_{1,1} \theta_{2,1}$,
$\:\:\:\theta_{12,i}=\theta_{1,i} -\theta_{2,i}$,
$\:\:\:Z_{12}^{2}=-\theta_{1,2}\theta_{2,2}$. Using this last expression we
obtain $(Z_{12}^{2})^{m}=0 \bk \forall \; m\geq2$.
This leads to:
\bea
f(Z_{1}) \! &=& \! \sum_{n=0}^{\infty} \frac{1}{n!} (Z_{12}^{1})^{n}
\partial_{z_{2}}^{n} (1+Z_{12}^{2} \partial_{z_{2}}) (1+\theta_{12,1}
D_{2,1}) (1+\theta_{12,2} D_{2,2}) f(Z_{1}^{2}) \nn \\
\Rightarrow \bk\bk f(Z_{1}) &=& 
\sum_{n=0}^{\infty} \frac{1}{n!} [(Z_{12}^{1})^{n}
\partial_{z_{2}}^{n} + (Z_{12}^{1})^{n} Z_{12}^{2}
\partial_{z_{2}}^{n+1}] (1+\theta_{12,1}
D_{2,1}) (1+\theta_{12,2} D_{2,2}) f(Z_{1}^{2}) \nn \\
&=& \sum_{n=0}^{\infty} \frac{1}{n!}
[\underbrace{(Z_{12}^{1})^{n} + n (Z_{12}^{1})^{n-1} Z_{12}^{2}}_{(z_{1}
-z_{2}-\theta_{1,1}\theta_{2,1}-\theta_{1,2}\theta_{2,2})^{n}}]
\partial_{z_{2}}^{n} (1+\theta_{12,1}
D_{2,1}) (1+\theta_{12,2} D_{2,2}) f(Z_{1}^{2}) \; . \nn
\eea
Repeated application of this step until we reach 
$Z_{2}=Z_{1}^{N}$ completes the proof.
\eprf

We conclude this subsection with the main theorem of supercontour integration 
techniques: the Cauchy formulae. These formulae will be the essential
tools to evaluate commutation relations in superconformal field
theories.
\bth Cauchy formulae \label{th:cauchy}
\bea
\frac{1}{2 \pi i}\oint_{{\cal C}_{2}} dZ_{1}Z_{12}^{-n-1} f(Z_{1})&=&
\frac{1}{n!} \partial_{z_{2}}^{n} D_{2,1} \ldots D_{2,N} f(Z_{2}) \com \nn
\\ 
 \frac{1}{2 \pi i} \oint_{{\cal C}_{2}} dZ_{1} \theta_{12,i_{1}} \ldots
\theta_{12,i_{k}} Z_{12}^{-n-1} f(Z_{1})&=& 
\frac{(-1)^{Nk-\frac{(k+1)k}{2}}}{n!}
\epsilon_{i_{1}, \ldots, i_{k},(j_{1}),
\ldots,(j_{N-k})}
\partial_{z_{2}}^{n} D_{2,(j_{1})} \ldots \nn \\
&& \ldots D_{2,(j_{N-k})}
f(Z_{2}) \com \nn 
\eea
where the $j$'s are taken out of the complement of the $i$'s in $1, \ldots ,N$
in increasing order\footnote{Note that the function $f(Z_{1})$ 
has to be on the right of the integral.}. 
$\epsilon_{i_{1}, \ldots, i_{N}}$ denotes the totally
antisymmetric
tensor with $\epsilon_{1,\ldots,N}=1$.
Two particular cases of the last equation are:
\bea
\frac{1}{2\pi i}\oint_{{\cal C}_{2}} dZ_{1} \theta_{12,1} \ldots
\check{\theta}_{12,l} \ldots \theta_{12,N} Z_{12}^{-n-1} f(Z_{1}) &=&
\frac{1}{n!} (-1)^{N-l} \epsilon^{N} \partial_{z_{2}}^{n} D_{2,l} f(Z_{2})
\com \nn \\
\frac{1}{2\pi i}\oint_{{\cal C}_{2}} dZ_{1} \theta_{12,1} \ldots
 \theta_{12,N} Z_{12}^{-n-1} f(Z_{1})&=&
\frac{1}{n!} \epsilon^{N} \partial_{z_{2}}^{n}  f(Z_{2}) \pkt \nn
\eea
\eth
\bprf
Using the definition \refoth{\ref{def:scontint}} it is easy to show that
\bea
\frac{1}{2\pi i}\oint_{{\cal C}_{2}} \dif Z_{1} \theta_{12,1}
\ldots \theta_{12,N} Z_{12}^{-n-1} &=& \epsilon^{N} \delta_{n,0} \com
\eea
and the integral vanishes whenever the product
$\theta_{12,1}\ldots \theta_{12,N}$ is not complete.
We apply the super Taylor expansion [\eq{\ref{eq:taylor}}] to the integral
\bea
& \frac{1}{2\pi i}{\displaystyle \oint_{{\cal C}_{2}}}
 \dif Z_{1} \theta_{12,i_{1}} \ldots
\theta_{12,i_{k}} f(Z_{1}) Z_{12}^{-n-1} \com \label{eq:oint} 
\eea
where we expand $f(Z_{1})$ about $Z_2$.
The only contributions can arise from the term leading to a complete product
of $\theta_{12,1} \ldots \theta_{12,N}$.
Hence, for each $\theta_{12,j}$ missing in \eq{\ref{eq:oint}} we introduce a
derivative $D_{2,j}$ acting on $f(Z_{2})$. Finally, we use the usual Cauchy
formulae for contour integrals in the complex plane to obtain 
$\partial_{z_{2}}^{n}f(Z_{2})$. 
\eprf


\section{Superconformal field theory}

So far, we have defined the underlying geometry of the quantum field theory
which we want to construct. The theory is meant to be superconformally
invariant. Hence, we have a chiral stress-energy tensor $T(Z)$ generating the
local symmetry group of superconformal transformations.
The main objects in a conformal field theory are the {\it primary fields};
fields which play the r\^ole of the tensors. This means that they form 
conformally invariant differential forms. Exactly in the same way we
define the {\it (uncharged) superprimary fields}.
\bdf Uncharged superprimary fields\\
Fields $\Phi_{h}(Z)$ defined on a superconformal manifold 
$\pmb{${\cal S}$}_{N}$ transforming under
a superconformal transformation $Z \mapsto \bar{Z}$ as
\bea
\Phi_{h}(Z) = \bar{\Phi}_{h} ( \bar{Z} ) {(\det  D
\bar{\theta} )}^{\frac{2h}{N}} \com \label{eq:sprim}
\eea
are called (uncharged) superprimary fields. The complex number $h$ is called
the (super)conformal weight of $\Phi_{h}(Z)$.
This definition is chosen in such a way that the
differential form $[\Phi_{h}(Z)]^{\frac{N}{2h}} \dif Z$ is invariant.
\edf

We perform an infinitesimal superconformal transformation 
$z\mapsto \bar{z}=z+\delta z$ and $\theta_{i} \mapsto
\bar{\theta_{i}}=\theta_{i}+\delta \theta_{i}$ for $i=1,\ldots ,N$.
We define $\delta \Phi_{h}(Z)$ 
by $\Phi_{h}(Z) = \bar{\Phi}_{h}(\bar{Z})+\delta 
\Phi_{h}$. Hence
using \eq{\ref{eq:sprim}} we obtain $\delta \Phi_{h} =
\Phi_{h} (\bar{Z}) -\Phi_{h}(Z) ( \det D\bar{\theta} )^{\frac{-2h}{N}}$.
Since $( \det D\bar{\theta} )^{2} = \det D\bar{\theta}
(D\bar{\theta})^{T} $ we calculate the variation $\delta D$ defined
as $D \bar{\theta} (D\bar{\theta})^{T} =\bbbone + \delta D$.
For the trivial variation we have $\delta D=0$. 
Hence the variation of the determinant can be found as
$\delta (\det D \bar{\theta})^{2} = \trace{\delta D}$.
For the $i$-th diagonal element $(D_{k}\bar{\theta}_{(i)})
(D_{k}\bar{\theta}_{(i)})$ we obtain a variation of
$2 D_{(i)} \delta \theta_{(i)}$. Thus
$\delta (\det D\bar{\theta})^{2}=2D_{i} \delta \theta_{i}$. 
We use the Taylor expansion \eqoth{\ref{eq:taylor}} in order to calculate
the variation of $\Phi_{h} (\bar{Z})=
\Phi_{h}(Z) +\delta\theta_{i} D_{i} \Phi_{h}(Z) + \nu(Z) \partial_{z}
\Phi_{h}(Z)$ up to first order. 
Here $\nu(Z)=\delta z + \theta_{i}\delta\theta_{i}$ is the
infinitesimal version of the differential $\kappa$.
Taking these results together we reach
$\delta \Phi_{h}= [\delta \theta_{i} D_{i} + \nu(Z) \partial_{z} 
+\frac{2h}{N} (D_{i} \delta \theta_{i})] \Phi_{h}(Z)$. Finally, we want 
to write the variations $\delta \theta_{i}$ in terms of derivatives,
for which we use the definition of superconformal transformations
\eqoth{\ref{eq:supconf}}. Neglecting higher order terms in
$D_{j} \bar{z} = \bar{\theta}_{i} D_{j} \bar{\theta}_{i}$ leads to
$\delta\theta_{i} = D_{j} \delta z -\theta_{i} D_{j} \delta
\theta_{i}$ for $i=1,\ldots , N$. We replace then $\delta z$ by $\nu(Z)$:
$\delta\theta_{j} =\frac{1}{2}D_{j} \nu(Z)$ and
$D_{j} \delta\theta_{j} =\frac{N}{2} \partial_{z} \nu(Z)$.
We can thus give the infinitesimal transformation of
$\Phi_{h}(Z)$ under infinitesimal superconformal transformations:
\bth
Under an infinitesimal superconformal transformation $z \mapsto
\bar{z} = z + \delta z$ , $\theta_{i} \mapsto
\bar{\theta}_{i} = \theta_{i} + \delta \theta_{i}$ , the change of
an (uncharged) superprimary field is given by
\bea
\delta \Phi_{h} (Z) &=& \frac{1}{2} (D_{j} \nu (Z)) D_{j} \Phi_{h} (Z) + \nu (Z)
\partial_{z} \Phi_{h} (Z) + h (\partial_{z} \nu (Z)) \Phi_{h} (Z) 
\com \label{eq:spriminf}
\eea
where $\nu (Z)$ is the infinitesimal version of $\kappa$: $\nu (Z)
= \delta z + \theta_{i} \delta \theta_{i}$ and it corresponds to the
superdifference: $\bar{z}-z-\bar{\theta}_{i}\theta_{i}=\nu (Z)$.
\eth

The field theory we constructed so far contains the stress-energy tensor,
the superprimary fields and all the descendant fields obtained from the
superprimary fields by applying superconformal transformations, that
is acting with modes of $T(Z)$ on them. Altogether this forms the closed set 
$\bPhi_{N}$ of fields contained in the theory. The radial
quantisation procedure defines the meromorphic function 
$\phi_{1}(Z_{1}) \phi_{2}(Z_{2})$ of two fields
in $\bPhi_{N}$, which is meant to be understood
inside correlation functions such as $\bra{0}\phi_{1}(Z_{1}) 
\phi_{2}(Z_{2})\ket{0}$ for time-ordered points $Z_{1}$ and $Z_{2}$:
$\abs z_{1} \abs > \abs z_{2} \abs$. For $\abs z_{2} \abs >
 \abs z_{1} \abs$ we define $\phi_{1}(Z_{1}) \phi_{2}(Z_{2})$ to be
its analytic continuation. $\phi_{1}(Z_{1}) \phi_{2}(Z_{2})$ is called the
{\it operator product} of $\phi_{1}(Z_{1})$ and $\phi_{2}(Z_{2})$.
In these terms integrals over 
equal time commutators become contour integrals which
we extended to supercontour integrals.

We assumed that the superprimary fields and its descendants form the complete set
of fields $\bPhi_{N}$ for the theory. Hence, the function
$\Phi_{h_{1}}(Z_{1})\Phi_{h_{2}}(Z_{2})$ can be expanded 
about the superpoint $Z_{j}$ where the expansion coefficients
are superprimary fields or descendants of superprimary fields. This expansion is
called the {\it operator product expansion (OPE)} of the fields
$\Phi_{h_{1}}(Z_{1})$ and $\Phi_{h_{2}}(Z_{2})$. 

In a radially quantised theory 
the time-ordered Euclidean symmetry generator generating the infinitesimal
transformation $z\mapsto z+\delta z$ and $\theta_{i}\mapsto \theta_{i}
+ \delta \theta_{i}$ becomes
the contour integral $\frac{1}{2\pi i}\oint \dif Z\: \nu(Z) T(Z)$.
Here we have chosen $T(Z)$ in such a way that $\frac{1}{2\pi i}
\oint \dif Z \: T(Z)$ 
generates the infinitesimal change of $z$ and 
correspondingly $\frac{-1}{2\pi i}\oint \dif Z \: \theta_{i} T(Z)$ 
of $\theta_{i}$.
We use the infinitesimal transformation \eqoth{\ref{eq:spriminf}}
of the superprimary field
$\Phi_{h}(Z)$ with conformal weight $h$ in order to determine the singular 
terms of the OPE $T(Z_{1})\Phi_{h}(Z_{2})$:
\bea
\delta_{\nu} \Phi_{h}(W)=\frac{1}{2\pi i}\oint_{{\cal C}_{W}} \dif 
Z\nu(Z)T(Z)\Phi_{h}(W) 
\eea
\bea
\Rightarrow \bk 
T(Z_{1})\Phi_{h}(Z_{2})=\frac{h\pi_{12}^{N}}{Z_{12}^{2}}
\Phi_{h}(Z_{2})+\frac{1}{2} \frac{\Delta_{12,j}^{N}}{Z_{12}}
D_{2,j}\Phi_{h}(Z_{2}) +
\frac{\pi_{12}^{N}}{Z_{12}}\partial_{z_{2}}\Phi_{h}(Z_{2}) 
+ \ldots (reg) \com
\label{eq:tpope}
\eea
\bea 
\pi_{12}^{N}=\epsilon^{N} \theta_{12,1} \ldots \theta_{12,N} & , & 
\bk \Delta_{12,j}^{N}=\epsilon^{N} (-1)^{N-j} \theta_{12,1} \ldots 
\check{\theta}_{12,j} \ldots \theta_{12,N} \pkt \nn
\eea
\noi Here $\ldots (reg)$ indicates non-singular terms.

We can determine\footnote{This calculation is straightforward but not trivial.
For the case of $N=2$ see for instance Ref. \icite{blumenhagen}.}
most of the singular terms of the operator product
$T(Z_{1})T(Z_{2})$ by
performing two successive superconformal transformations 
$\delta_{\nu_{2}}\delta_{\nu_{1}}\Phi_{h}$ and evaluating the
double supercontour integral in two different ways using 
contour deformation. However, the singular terms are
fixed except for a term of the form $\frac{1}{Z_{12}^{4-N}}$;
and this is the only degree of freedom we can find. This term is
called the {\it central extension} term. It does not contribute
towards the infinitesimal change of $\Phi_{h}(W)$ and may
therefore be contained in the quantum field theory:
\bea
T(Z_{1})T(Z_{2}) \!\! &=&\!\! \frac{\hat{C}}{Z_{12}^{4-N}} + \frac{4-N}{2}
\frac{\pi_{12}^{N}}{Z_{12}^{2}} T(Z_{2})+
\frac{\Delta_{12,j}^{N} D_{2,j}}{2\: Z_{12}} T(Z_{2}) +
\frac{\pi_{12}^{N}}{Z_{12}} \partial_{z_{2}} T(Z_{2}) +\ldots (reg) \; .
\label{eq:ttope}
\eea

It is worth remarking that \eq{\ref{eq:ttope}} does not allow a central
extension for theories with $N \geq 4$ because the central extension
term does not belong to a singularity any more and hence will not
contribute to the commutation relations of the modes as we shall see.
For $0\leq N \leq 3$ we set $\hat{C}=\epsilon^{N} \frac{(3-N)!}{12} C$ which
will lead us to the central term of the Virasoro algebra.

We have now defined the main objects in a superconformal field theory. 
In order to look at the space of states of the physical model we
expand $T(Z)$ in its modes. Therefore we take the sequences $(i_{1},
\ldots,i_{I})$ where $i_{j}\in \{1,\ldots, N\}$ and define the 
operators:
\bea
J_{a}^{i_{1},\ldots,i_{I}}(Z_{2}) &=& \frac{1}{2\pi i}
\oint_{{\cal C}_{2}} \dif Z_{1} \theta_{12,i_{1}}
\ldots \theta_{12,i_{I}} Z_{12}^{a+1-\frac{I}{2}} T(Z_{1}) \pkt
\label{eq:sgenq}
\eea
The index $a$ is chosen out of $\bbbz$ if $I$ is even in which case
$J_{a}^{i_{1},\ldots,i_{I}}(Z_{2})$ is classified as even or otherwise
$a$ is taken out of $\bbbzh$ and $J_{a}^{i_{1},\ldots,i_{I}}(Z_{2})$ is
qualified as odd. $J_{a}^{i_{1},\ldots,i_{I}}(Z_{2})$
is trivial if $(i_{1},\ldots,i_{I})$ contains the same number twice.
Moreover if $(i_{1}^{\prime},\ldots,i_{I}^{\prime})$ is 
a reordering of $(i_{1},\ldots,i_{I})$, then the corresponding operators differ
at most by a sign factor. 
The mode $J_{a}^{i_{1},\ldots,i_{I}}(Z_{2})$ corresponds to the 
expansion term
$\epsilon^{N} \epsilon_{i_{1},\ldots,i_{I},
(j_{1}),\ldots,(j_{N-I})}\theta_{12,(j_{1})}\ldots \theta_{12,(j_{N-I})}
Z_{12}^{-a-2+I/2}$ of $T(Z_1)$
where the $j_{k}$'s are taken from the complement
of $(i_{1},\ldots,i_{I})$ in the sense of a set.
\eq{\ref{eq:tpope}} allows us to derive the 
commutation relations of the modes. It is common practice to consider the
symmetry algebra generators taken at the origin. This is not a
constraint since we can shift the generators to any other point 
by superconformal conjugation. Nevertheless, the commutation relations
do not depend on the chosen base point. Instead of  
$J_{a}^{i_{1},\ldots,i_{I}}(0)$ we shall just write $J_{a}^{i_{1},\ldots,i_{I}}$
unless we explicitly want to base the generator on a different point than the
origin in which case we write $J_{a}^{i_{1},\ldots,i_{I}}(Z)$.
Standard contour deformation techniques result in
\bea
[J_{a}^{i_{1},\ldots,i_{I}},J_{b}^{j_{1},\ldots,j_{J}}]_{S} &=&
\oint_{{\cal C}_{0}} \frac{\dif Z_{2}}{2\pi i}
\oint_{{\cal C}_{2}} \frac{\dif Z_{1}}{2\pi i} \theta_{10,i_{1}}
\ldots \theta_{10,i_{I}} \nn \\ 
&& \theta_{20,j_{1}} \ldots
\theta_{20,j_{J}} Z_{10}^{a+1-\frac{I}{2}} Z_{20}^{b+1-\frac{J}{2}}
T(Z_{1})T(Z_{2}) \pkt
\eea
Performing\footnote{By convention we moved $T(Z_{1})T(Z_{2})$ to
the right of the integral.} the supercontour integrals
produces the following result\footnote{The falling product $\fall{x}{n}$
is defined as $x(x-1)\ldots(x-n+1)$ for $n\in\bbbn$ and $\fall{x}{0}=1$.}:
\bea
[J_{a}^{i_{1},\ldots,i_{I}},J_{b}^{j_{1},\ldots,j_{J}}]_{S} &=&
(-1)^{N(I+J)} [a(1-\frac{J}{2})-b(1-\frac{I}{2})] \: 
J_{a+b}^{i_{1},\ldots,i_{I},j_{1},\ldots,j_{J}} \nn \\
&& \!\!\!\!\!\!\!\! 
+ \frac{(-1)^{N(I+J)}}{2} \sum_{p=1}^{I} \sum_{q=1}^{J}
(-1)^{I+p+q} \delta_{i_{p},j_{q}} 
J_{a+b}^{i_{1},\ldots ,\check{i}_{p},\ldots,
i_{I},j_{1},\ldots,\check{j}_{q},\ldots,j_{J}} \\
&& \!\!\!\!\!\!\!\!
+\frac{C\fall{a+1-\frac{I}{2}}{3-I}}{12} 
\delta_{(j_{1}),\ldots,(j_{J})}^{(i_{1}),\ldots,(i_{I})}
\delta_{a+b,0}
(-1)^{NI-\frac{(I+1)I}{2}}
\epsilon_{(i_{1}),\ldots,(i_{I})} 
\epsilon_{(j_{1}),\ldots,(j_{J})} 
\delta^{\{0,1,2,3\}}_{N} , \nn
\eea
where\footnote{If $(i_{1},\ldots,i_{I})$ is the empty set we define
$\epsilon_{\emptyset}=1$. However, $\epsilon_{\emptyset}$ should
not be confused with $\epsilon^{N}$.}
$\delta^{S}_{N}=1$ if $N\in S$ or otherwise $\delta^{S}_{N}=0$
and $\delta_{j_{1},\ldots,j_{J}}^{i_{1},\ldots,i_{I}}=1$
if $(i_{1},\ldots,i_{I})$ and $(j_{1},\ldots,j_{J})$ are the same up to
reordering or otherwise
$\delta_{j_{1},\ldots,j_{J}}^{i_{1},\ldots,i_{I}}=0$.
For the generators $J_{m}$ with $I=0$ we easily obtain the commutation 
relations
\bea
[J_{m},J_{n}] &=& (m-n) J_{m+n} +\frac{C}{12}(m^{3}-m) \delta_{m+n,0}
\delta^{\{0,1,2,3\}}_{N} \: .
\eea
Hence\footnote{Note that $\hat{C}$ commutes with
all generators $J_{a}^{i_{1},\ldots,i_{I}}$.}
we find the Virasoro algebra as a subalgebra of the symmetry algebra 
generated by the modes \eqoth{\ref{eq:sgenq}}. 
This algebra is called the {\it Neveu-Schwarz superconformal
algebra with parameter $N$} which mathematicians denote by $\scn{N}$.
As expected \eq{\ref{eq:scom}} is a representation\footnote{We still have to
scale the generators with factors of the complex unit
$i$ in order to obtain the notation of \eq{\ref{eq:scom}}.}
of $\scn{N}$ with $\hat{C}=0$.
In the following two sections we give the results
for $N=1$ and $N=2$ explicitly.
We have derived the algebras $\scn{N}$ by using superfield formalism.
Independently to this approach one could try to find superconformal extensions
of the Virasoro algebra just on Lie superalgebra level. This
could be done without constructing 
the underlying superconformal geometry in terms of extended differential 
manifolds. The geometry would then be
defined thanks to the generators. 

\noi To conclude this section we calculate the commutators of $\scn{N}$ generators
with (uncharged) superprimary fields:
\bea
[J_{a}^{i_{1},\ldots,i_{I}}(Z_{0}),\Phi_{h}(Z_{2})] \!\! &=& \!\!
\oint_{{\cal C}_{2}}\frac{1}{2\pi i} \theta_{10,i_{1}} \ldots 
\theta_{10,i_{I}} Z_{10}^{a+1-\frac{I}{2}} T(Z_{1})\Phi_{h}(Z_{2}) \pkt
\nn 
\eea
We use \eq{\ref{eq:tpope}} to evaluate the contour integrals. Hence
we can write the commutator as a differential operator acting on the 
superprimary field:
\bth
The commutator of the algebra generator 
$J_{a}^{i_{1},\ldots,i_{I}}(Z_{0})$ with the superprimary field 
$\Phi_{h}(Z_{2})$ can be written as
\bea
[J_{a}^{i_{1},\ldots,i_{I}}(Z_{0}),\Phi_{h}(Z_{2})] &=&
(-1)^{NI} \Bigl[ h(a+1-\frac{I}{2})\theta_{20,i_{1}} \ldots
\theta_{20,i_{I}} Z_{20}^{a-\frac{I}{2}} 
+ \theta_{20,i_{1}} \ldots
\theta_{20,i_{I}} Z_{20}^{a+1-\frac{I}{2}} \partial_{z_{2}} \nn \\
&&
-\frac{(-1)^{I}}{2} \sum_{p=1}^{I} (-1)^{p}
\theta_{20,i_{1}} \ldots \check{\theta}_{20,i_{p}} \ldots
\theta_{20,i_{I}} Z_{20}^{a+1-\frac{I}{2}} D_{2,i_{p}} \label{eq:primcom}
\\
&&
+\frac{(a+1-\frac{I}{2}) (-1)^{I}}{2} 
\theta_{20,i_{1}} \ldots 
\theta_{20,i_{I}} \theta_{20,j} 
Z_{20}^{a-\frac{I}{2}} D_{2,j} \Bigr] \Phi_{h}(Z_{2}) \pkt \nn
\eea
\eth
\noi In particular we obtain for the Virasoro generators $L_{m}=J_{m}$:
\bea
[L_{m},\Phi_{h}(Z)]&=& \Bigl[ h(m+1) z^{m} + z^{m}\partial_{z}
+\half (m+1) \theta_{j} z^{m} \partial_{\theta_{j}} \Bigr]
\Phi_{h}(Z) \pkt
\eea


\section{$N=1$ Superconformal theories}

The simplest superconformal extensions of conformal field theories are 
the $N=1$ superconformal field theories\cite{friedan}. 
The embedding structure of the corresponding
highest weight representations has been analysed by
Astashkevich\cite{ast}.
Their superconformal
structure is not large enough to show significant differences to the
representation theory of the Virasoro algebra. This makes the study of
$N=1$ superconformal representation theory
not very spectacular. 
Maybe this is the reason
why the same was suspected for the representation theory of superconformal
theories with bigger $N$ and hence literature did not treat
the $N=2$ representation
theory as one could wish. However, we showed in 
reference \icite{paper2} why
especially $N=2$ representations are much more appealing and their 
structures much different than already discussed by other 
authors\cite{dobrev,kiritsis2}.
Besides and as an exercise we want to use the definitions from the
previous sections to define explicitly $N=1$ superconformal theories.
We define a quantum field theory containing a chiral stress-energy
tensor $T(Z)$, which generates the local $N=1$ superconformal
transformations, and we have superprimary fields $\Phi_{h}(Z)$ which
have in our quantisation scheme according to \eq{\ref{eq:tpope}}
an operator product expansion with the stress-energy tensor of the form
\bea
T(Z_{1})\Phi_{h}(Z_{2})&=& \frac{h\theta_{12}}{Z_{12}^{2}} \Phi_{h}
(Z_{2})+\half \frac{1}{Z_{12}} D_{2}\Phi_{h}(Z_{2})+
\frac{\theta_{12}}{Z_{12}}\partial_{z_{2}} \Phi_{h}(Z_{2})
+ \ldots (reg) \pkt \label{eq:eq:N1tpope}
\eea 
We obtain the symmetry algebra of the
theory by expanding the stress-energy tensor:
\bea
T(Z_{1})   
&=& \theta_{12} \sum_{m\in\bbbz} Z_{12}^{-m-2}L_{m}(Z_{2})
+\sum_{r\in \bbbz_{\frac{1}{2}}} Z_{12}^{-r-\frac{3}{2}} G_{r}(Z_{2}) 
\com \label{eq:N1texp} 
\eea
\bea
L_{m}(Z_{2}) & = J_m(Z_2)
=&\oint_{{\cal C}_{2}} \frac{\dif Z_{1}}{2\pi i} Z_{12}^{m+1}
T(Z_{1}) \com \nn \\
G_{r}(Z_{2})& = J_r^1(Z_2)
=&\oint_{{\cal C}_{2}}  
\frac{\dif Z_{1}}{2\pi i} \theta_{12} Z_{12}^{r+\half}
T(Z_{1}) \com \label{eq:N1algc}
\eea
where ${\cal C}_{2}$ is a $N=1$ supercontour about $Z_{2}$.
Finally we need to know the operator product of $T(Z)$ with itself.
According to \eq{\ref{eq:ttope}} we obtain:
\bea
T(Z_{1})T(Z_{2}) &=& \frac{C}{6\: Z_{12}^{3}} +\frac{3}{2}
\frac{\theta_{12}}{Z_{12}^{2}} T(Z_{2}) +\half \frac{1}{Z_{12}}
D_{2} T(Z_{2}) +\frac{\theta_{12}}{Z_{12}} \partial_{z_{2}}
T(Z_{2})+ \ldots (reg) \pkt \label{eq:N1ttope}
\eea
This enables us to derive the commutators of the symmetry algebra
generators [\eqs{\ref{eq:N1algc}}] using \eq{\ref{eq:N1ttope}} and standard
contour deformation methods. The well-known
result contains the Virasoro algebra and the
generators of one anticommuting field:
\bdf
The $N=1$ Neveu-Schwarz superconformal algebra $\scn{1}$ has the following
commutation relations:
\bea
[L_{m},L_{n}] & = & (m-n)L_{m+n} + \frac{C}{12} (m^{3}-m)
\delta_{m+n,0} \com \nn \\
\ [L_{m},G_{r}] &=& (\frac{m}{2}-r) G_{m+r} \com \label{eq:N1alg} \\
\{ G_{r},G_{s} \} &=& 2 L_{r+s} + \frac{C}{3}(r^{2}-\frac{1}{4} )
\delta_{r+s,0} \com \nn \\
\ [L_{m},C] &=& [G_{r},C] \; = \; 0 \com \nn
\eea
where $m,n\in\bbbz$ and $r,s\in\bbbz_{\frac{1}{2}}$.
\edf
We have found the algebra \eqoth{\ref{eq:N1alg}} by using superfield formalism.
As mentioned earlier there may be other super extensions of the Virasoro
algebra.
In the case of $N=1$ algebras one finds another $N=1$ superconformal algebra.
The only difference to \eqoth{\ref{eq:N1alg}} is that the operators $G_{r}$ have
integer indices rather than half integer indices. This algebra is commonly
called the {\it $N=1$ Ramond algebra}.


\section{$N=2$ Superconformal theories}

The algebra of chiral superconformal transformations in $N=2$ superconformal
space\cite{cohn} is generated
by the super stress-energy tensor $T(Z_{1})$, where $Z_{1}$
denotes a superpoint $(\! z_{1},\theta_{1,1},\theta_{1,2}\! )$
and $D_{2,1}$ the superderivative $\pder{}{\theta_{2,1}}
+\theta_{2,1}\pder{}{z_{2}}$. Superconformal
invariance of the theory determines the singular part of the
operator product of $T(Z)$ with itself, according to \eqoth{\ref{eq:ttope}}.
In this case the central non-fixed term is of the form 
$\frac{1}{Z_{12}^{2}}$:
\bea
T(Z_{1}) \: T(Z_{2}) &=& -\frac{C}{12 Z_{12}^{2}} + \left( -
\frac{\theta_{12,1} \theta_{12,2}}{Z_{12}^{2}} +  
\frac{\theta_{12,2} D_{2,1} - \theta_{12,1} D_{2,2}}{2\: Z_{12}} 
 - \frac{\theta_{12,1} \theta_{12,2}}{Z_{12}} \partial_{z_{2}} 
\right) T(Z_{2}) \nn \\ && + \ldots (reg) \; . \label{eq:ttn2}
\eea

\noi 
\eq{\ref{eq:ttn2}} agrees with the result for the OPE derived by 
Blumenhagen\cite{blumenhagen}.
Expanding $T(Z)$ in its modes allows us to find the symmetry algebra
generators\footnote{As mentioned earlier $L_{m}(0)$ is 
denoted by $L_{m}$ and respectively $T_{m}(0)$ by $T_{m}$ etc.}: 
\bea
T(Z_{1}) & = & -\theta_{12,1} \theta_{12,2} \sum_{m \in \bbbz} Z_{12}^{-m-2}
L_{m}(Z_{2})-\frac{1}{2} \sum_{r \in \bbbzh}
Z_{12}^{-r-\frac{3}{2}} \theta_{12,2} G_{r}^{1}(Z_{2}) \nn \\
&& +\frac{1}{2} \sum_{r \in \bbbzh}
Z_{12}^{-r-\frac{3}{2}} \theta_{12,1} G_{r}^{2}(Z_{2}) - \frac{1}{2}  
i \sum_{m \in \bbbz} Z_{12}^{-m-1}T_{m}(Z_{2}) \com \label{eq:texp1} 
\eea
\bea
L_{m}(Z_{2}) &  = J_m(Z_2) = & \oint_{{\cal C}_{2}} 
\frac{dZ_{1}}{2\pi i} Z_{12}^{m+1} T(Z_{1})
\com \nn \\
G_{r}^{i}(Z_{2}) &  = 2J_r^i(Z_2) = & 2 \oint_{{\cal C}_{2}}
\frac{dZ_{1}}{2\pi i} \theta_{12,i}
Z_{12}^{r+\frac{1}{2}}T(Z_{1})  \com \label{eq:gen}\label{eq:texp2} \\
T_{m}(Z_{2}) &  = 2iJ_m^{21}(Z_2) = & -2i \oint_{{\cal C}_{2}}
\frac{dZ_{1}}{2\pi i} \theta_{12,1}
\theta_{12,2} Z_{12}^{m} T(Z_{1}) \pkt \nn 
\eea
If we use complex coordinates for the odd generators
\bea
 \theta^{\pm} &=& \frac{1}{\sqrt{2}} (\theta_{1} \pm i \theta_{2}) \com 
\label{eq:compl1} \\ 
G^{\pm} &=& \frac{1}{\sqrt{2}} (G^{1} \pm i G^{2}) \com \\
 D^{\pm}& =& \frac{1}{\sqrt{2}} (D_{1} \pm i D_{2}) =  
\frac{\partial}{\partial \theta^{\mp}} + \theta^{\pm}
\frac{\partial}{\partial z} \com \label{eq:compl3}
\eea
we find for the symmetry algebra $\sct$ a
decomposition in the Virasoro algebra, two anticommuting
fields, and a
$\rm{U}(1)$ Kac-Moody algebra with the commutation relations
\bea
[L_{m},L_{n}] & = & (m-n) L_{m+n} + \frac{C}{12} \:(m^{3}-m)\:
\delta_{m+n,0} \com \nn \\
\ [L_{m},G_{r}^{\pm}] & = & (\frac{1}{2} m-r) G_{m+r}^{\pm} \com \nn \\
\ [L_{m},T_{n}] & = & -n T_{m+n} \com \nn \\
\ [T_{m},T_{n}] & = & \frac{1}{3} C m \delta_{m+n,0}  
\com \label{eq:cr} \\
\ [T_{m},G_{r}^{\pm}] & = & \pm G_{m+r}^{\pm} \com \nn \\
\ \{ G_{r}^{+},G_{s}^{-}\} & = & 2 L_{r+s}+(r-s) T_{r+s} +\frac{C}{3}
(r^{2}-\frac{1}{4}) \delta_{r+s,0} \com \nn \\
\ [L_{m},C] & = & [T_{m},C] \;\; = \;\; [G_{r}^{\pm},C] 
\;\;=\;\; 0 \com \nn \\
\ \{G_{r}^{+},G_{s}^{+}\} & = & \{G_{r}^{-},G_{s}^{-}\}=0 , \;\;\;\;
\;\;\;\; m,n \in \bbbz ; \;\; r,s \in \bbbz_{\frac{1}{2}} \pkt \nn 
\eea
\eq{\ref{eq:tpope}} gives us the singular terms of the 
operator product of $T(Z_{1})$ and
superprimary fields $\Phi_{h} (Z)$. It turns out, that with respect to
the adjoint representation, $\Phi_{h}$ has the $T_{0}$ eigenvalue $0$.
This restricts the highest weight representations of
\eqoth{\ref{eq:cr}} as we want $T_{0}$ in the Cartan subalgebra of $\sct$.
We can extend the theory by introducing {\it
charged superprimary fields}
$\Phi_{h,q}(Z)$. The action of superconformal
transformations on $\Phi_{h,q}(Z)$ is altered by a term 
$\frac{q}{2 Z_{12}}$ in the operator product expansion of
$T(Z_{1})\Phi_{h,q}(Z_{2})$. This corresponds to 
a standard $U(1)$ charge term\footnote{This term was not included in
\eq{\ref{eq:tpope}} since the number of charges increases with $N$.}.
\bdf
We define the charged superprimary fields on the superconformal space
$\pmb{${\cal S}$}_{2}$ by the operator product expansion
\bea
T(Z_{1}) \Phi_{h,q} (Z_{2}) & = & - \frac{h \theta_{12,1}
\theta_{12,2}}{Z_{12}^{2}} \Phi_{h,q}(Z_{2})+\frac{1}{2} 
\frac{\theta_{12,2} D_{2,1} - \theta_{12,1} D_{2,2}}{Z_{12}}
\Phi_{h,q}(Z_{2}) \nn \\
&& -\frac{\theta_{12,1} \theta_{12,2}}{Z_{12}}
\partial_{z_{2}} \Phi_{h,q}(Z_{2}) - \frac{q}{2 Z_{12}} i \Phi_{h,q} (Z_{2})
\ldots \pkt
\eea
We call $h$ the conformal weight of $\Phi_{h,q}$ and $q$ its 
conformal charge, corresponding to the scaling dimensions of $L_{0}$
and $T_{0}$ transformations respectively. 
\edf
Using \eqs{\ref{eq:texp2}} and performing the contour integrals we
find the infinitesimal transformations for all generators:
\bea
\ [L_{m},\Phi_{h,q} (Z)] &=& \Bigl[ h(m+1) z^{m} + \frac{1}{2} (m+1) z^{m}
(\theta^{+} D^{-} + \theta^{-} D^{+}) +z^{m+1} \partial_{z} \nn \\
&& +\frac{q}{2} \theta^{+} \theta^{-} z^{m-1} m (m+1) \Bigr]
\Phi_{h,q}(Z) \com \nn \\
\ [G_{r}^{\pm},\Phi_{h,q} (Z)] &=& \Bigl[ 2 h (r+\frac{1}{2}) \theta^{\pm}
z^{r-\frac{1}{2}} -z^{r+\frac{1}{2}} D^{\pm} \pm \theta^{+} \theta^{-}
(r+\frac{1}{2}) z^{r-\frac{1}{2}} D^{\pm} \nn \\
&& + 2 \theta^{\pm} z^{r+\frac{1}{2}} \partial_{z} \pm q
\theta^{\pm} z^{r-\frac{1}{2}} (r+\frac{1}{2}) \Bigr] \Phi_{h,q}(Z) \com 
\label{eq:phi_cr} \\
\ [T_{m},\Phi_{h,q}(Z)] &=& \Bigl[ 2 h \theta^{+} \theta^{-} m z^{m-1} +
z^{m} (\theta^{-} D^{+} - \theta^{+} D^{-}) + 2 \theta^{+} \theta^{-}
z^{m} \partial_{z} \nn \\
&& + q z^{m} \Bigr] \Phi_{h,q}(Z) \pkt\nn 
\eea
Once more, we note that the $N=2$ superconformal algebra 
$\sct$ which we consider is known as the $N=2$ {\it Neveu-Schwarz} 
or {\it antiperiodic} algebra.
Reference \icite{schwimmer} has shown that it is isomorphic
to the $N=2$ {\it Ramond}
or {\it periodic} algebra what makes a separate discussion redundant. 

We can write $\sct$ in a triangular 
decomposition\footnote{There are in fact triangular decompositions
having a four-dimensional space ${\cal H}_{2}$. However, these decompositions
are not consistent with our $\bbbz_{2}$ grading.}
such that ${\cal H}_{2}$
contains the energy operator $L_{0}$:
$\sct=\sct^{-} 
\oplus {\cal H}_{2} \oplus \sct^{+}$, where 
${\cal H}_{2}=\spn{L_{0},T_{0},C}$ is the
grading preserving Cartan subalgebra, and
\bea
\sct^{\pm}=
\spn{ L_{\pm n},T_{\pm n},
G_{\pm r}^{+},G_{\pm r}^{-}: n
\in \bbbn, r \in \bbbn_{\frac{1}{2}} }. \nn
\eea
A simultaneous eigenvector
$\ket{h,q,c}$ of ${\cal H}_{2}$ with $L_{0}$, $T_{0}$ and $C$
eigenvalues $h$,
$q$ and $c$ respectively
and vanishing $\sct^{+}$ action, $\sct^{+}
 \ket{h,q,c} =0$,
is called a {\it highest weight vector}. 
It is easy to see that a
primary field $\Phi_{h,q}$ generates a highest weight vector
$\ket{h,q,c}$ on the vacuum: $\ket{h,q,c}=\Phi_{h,q}(0) \ket{0}_{c}$.
$\ket{0}_{c}$ denotes the vacuum with fixed central extension $C\ket{0}_{c}
=c\ket{0}_{c}$.
The Verma module ${\cal V}_{h,q,c}$
is defined as the $\sct$ left-module $U(\sct)
 \otimes_{{\cal
H}_{2} \oplus \sct^{+}} \ket{h,q,c}$,
where $U(\sct)$ denotes the
universal enveloping algebra of $\sct$. 

The representation theory of the $N=2$ Neveu-Schwarz algebra contains 
many surprising features which have not appeared in any other conformal
field theory so far. Singular vectors can be degenerated\cite{paper2},
embedded modules may not be complete and subsingular vectors
can appear\cite{beatriz1,beatriz2,paper7}, 
rousing the curiosity about what one
has to deal with for even higher $N$.


\section{Transformation of primary fields}

As an application of the formalism developed above we compute
the global transformation rules for (uncharged) primary fields under
transformations generated by the algebra generators. Restricted to the
Virasoro case our results represent the 
inverse problem of the transformation
formula found by Gaberdiel\cite{matthias}. There the author 
derived for the Virasoro case the algebra element that 
belongs to a given holomorphic
coordinate transformation, whilst we are interested in the 
holomorphic coordinate transformation corresponding to the given
algebra generators. Furthermore, in our superfield framework 
we can very easily obtain transformation rules also for all 
the superconformal cases as we shall demonstrate
in some specific examples.

The only globally defined conformal transformations on the 
Riemann sphere are the M\"obius transformations. 
They are generated\cite{ginsparg} by $L_{-1}$, $L_{0}$ and $L_{1}$.
$L_{-1}$ and $L_{0}$  respectively 
correspond to translations and to scaling transformations
on the Riemann surface\footnote{Note that this is after mapping the system
from the cylinder to the complex plane 
for radial quantisation.}:
\bea
e^{\lambda L_{-1}} \Phi_{h}(z) e^{-\lambda L_{-1}} &=& \Phi_{h}(z+\lambda) 
\com \nn \\
\lambda^{L_{0}} \Phi_{h}(z) {\lambda}^{- L_{0}} &=& {\lambda}^{h} 
\Phi_{h}(\lambda  z) \pkt  
\eea
$\lambda L_{1}$ corresponds to the coordinate transformation
$z\mapsto\frac{z}{1-\lambda z}$. 
Since the central extension does not
contribute towards the infinitesimal change
of $\Phi_{h}$, we can obtain the transformation rules 
by using the action of the de Witt algebra,
which forms a representation of the Virasoro algebra with $C=0$. 
In this section we want to find general transformation 
formulae for the primary fields under transformations
generated by the Virasoro generators. 
We shall then extend this result to
the $N=1$ and $N=2$ superconformal algebra.
The extension to superconformal algebras with $N\geq3$ follows
the same rules as we shall see.

\subsection{Primary fields in the Virasoro case}
The action of the M\"obius generators $L_{-1},L_{0}$ and $L_{1}$ on a 
primary field $\Phi_{h}(z)$ is well-known 
in conformal field theory\cite{ginsparg}. 
In a first step we generalise these results to the Virasoro generators
$L_{n}$:
\bea
& e^{\lambda L_{n}} \Phi_{h} (z) e^{-\lambda L_{n}}  \pkt
\label{eq:transf}
\eea
\noi We use the identity\footnote{Successive commutators are defined
as $[A,B]_i=[A,[A,B]_{i-1}]$ and $_i[A,B]=[_{i-1}[A,B],B]$
with $[A,B]_0=B$ and $_0[A,B]=A$.} 
\bea
e^{A} \: B \: e^{-A}  &=&  \sum_{j=0}^{\infty}\frac{1}{j!}[A,B]_{j}
\com
\eea
to rewrite \eq{\ref{eq:transf}}:
\bea
e^{\lambda L_{n}} \Phi_{h} (z) e^{-\lambda L_{n}}
&=& \sum_{j=0}^{\infty} \frac{\lambda^{j}}{j!} [L_{n},\Phi_{h}(z)]_{j} \pkt
\label{eq:transsuc}
\eea
Substituting \eq{\ref{eq:vircomphi}}
in the successive commutator of \eq{\ref{eq:transsuc}} leads us to:
\bea
e^{\lambda L_{n}} \Phi_{h} (z) e^{-\lambda L_{n}}
&=& \sum_{j=0}^{\infty} \frac{\lambda^{j}}{j!}
[z^{n+1}\partial_{z}+h(n+1)z^{n}]^{j} \, \Phi_{h}(z) \nn \\
&=&
e^{\lambda[z^{n+1}\partial_{z}+h(n+1)z^{n}]} \, \Phi_{h}(z)
\pkt
\eea
Before we continue, we prove the following differential
operator identity which will constitute to be the main tool for the rest of
this section:
\bth \label{th:mir}
\bea
e^{\lambda[z^{n+1}\partial_{z}+h(n+1)z^{n}]} 
&=& \left[ { \left.\frac{\partial
(e^{\lambda w^{n+1}\partial_{w}} w)}{\partial w} \right|_{w=z}}\right]^{h}
e^{\lambda z^{n+1}\partial_{z}} \nn \\
&=& \left( \left. e^{\lambda \partial_{w}
w^{n+1}} \right|_{w=z} \right)^{h} 
e^{\lambda z^{n+1}\partial_{z}} \pkt
\label{eq:mir} 
\eea
\eth
\bprf
To prove the first part\footnote {We are grateful to Adrian Kent for
this proof.} of 
\eq{\ref{eq:mir}} it is sufficient to show it by
acting on the monomials $z^{m}$
for any positive integer m. 
By straightforward calculation we obtain the following identities
for integer $n\neq 0$:
\bea
 e^{\lambda z^{n+1} \partial_{z}}z&=&\frac{z}{(1-n\lambda
z^{n})^{\frac{1}{n}}} \com \label{eq:id1} \\
 e^{\lambda \partial_{z} z^{n+1}}&=&\frac{1}{(1-n\lambda
z^{n})^{\frac{n+1}{n}}} \pkt \label{eq:id2}
\eea
\noi For $n=0$ we have the identities:
\bea
 e^{\lambda z \partial_{z}}z&=&e^{\lambda}z \label{eq:id3} \com \\
 e^{\lambda \partial_{z} z}&=&e^{\lambda} \label{eq:id4} \pkt
\eea
Using \eq{\ref{eq:id2}}
and acting on the monomial $z^{m}$ with the first of the 
\eqs{\ref{eq:mir}} leads to:
\bea
 \sum_{j=0}^{\infty} \frac{\lambda^{j}}{j!} z^{m+jn} [m+h(n+1)]
[m+h(n+1)+n] \ldots [m+h(n+1)+(j-1)n] \nn \\
 = \frac{1}{(1-n\lambda
z^{n})^{\frac{n+1}{n}h}} \sum_{j=0}^{\infty} \frac{\lambda^{j}}{j!}
z^{m+jn} m [m+n] \ldots [m+(j-1)n] \pkt \nn
\eea 
We expand both sides in power series in $\lambda$ about $0$ and
compare the coefficients. We thus compare the
derivatives $\left. \frac{\partial^{q}}{\partial\lambda^{q}}
\right|_{\lambda =0}$ on both sides for any positive integer $q$. 
The left-hand side turns into\footnote{For $q=0$ the expression should 
be $z^{m}$.}:
\bea
z^{m+qn}[m+h(n+1)][m+h(n+1)+n] \ldots [m+h(n+1)+(q-1)n] \pkt \nn
\eea
In order to take the derivative of the right-hand side we use the
differentiation rule:
\bea
\frac{\partial^{q}}{\partial\lambda^{q}} A(\lambda ) B(\lambda ) =
\sum_{r=0}^{q} \left( \begin{array}{c} q \\ r \end{array} \right)
\left( \frac{\partial^{r} A(\lambda)}{\partial \lambda^{r}} \right)
\left( \frac{\partial^{q-r} B(\lambda)}{\partial \lambda^{q-r}} \right)
\bk . \nn
\eea
Hence, we obtain for the right-hand side:
\bea
\sum_{r=0}^{q} \left( \begin{array}{c} q \\ r \end{array} \right)
z^{m+qn} h(n+1)[h(n+1)+n] \ldots [h(n+1)+(r-1)n] \nn \\ m [m+n] \ldots
[m+(q-r-1)n] \pkt \nn
\eea
Therefore, we only need to prove the following identity:
\bea
& [m+h(n+1)] [m+h(n+1)+1] \ldots [m+h(n+1)+(q-1)n] \nn \\
& = {\displaystyle\sum_{r=0}^{q}} \left( \begin{array}{c} q \\ r \end{array} \right) m
[m+n] \ldots [m+(q-r-1)n] [h(n+1)] \ldots [h(n+1)+(r-1)n]. \label{eq:indh}
\eea
We show \eq{\ref{eq:indh}} by induction on $q$:
\eq{\ref{eq:indh}} is obviously valid for $q=1$.
We assume \eqoth{\ref{eq:indh}} is true for $q$ and we show that 
it is true for $q+1$:
\bea
&&[m+h(n+1)]\ldots [m+h(n+1)+qn] \nn \\
&&=\sum_{r=0}^{q} \left( \begin{array}{c} q \\ r \end{array} \right) m
\ldots [m+(q-r-1)n] [h(n+1)] \ldots [h(n+1)+(r-1)n]\nn \\ &&[m+h(n+1)+qn]
\nn \\
&&=\sum_{r=0}^{q} \left( \begin{array}{c} q \\ r \end{array} \right) m
\ldots [m+(q-r-1)n] [h(n+1)] \ldots [h(n+1)+(r-1)n]\nn \\ &&
[h(n+1)+rn+m+(q-r)n] \nn \\
&& =\sum_{r=1}^{q+1} \left( \begin{array}{c} q \\ r-1 \end{array}
\right) m \ldots [m+(q-r)m] [h(n+1)] \ldots [h(n+1)+(r-1)n] \nn \\
&& + \sum_{r=0}^{q} \left( \begin{array}{c} q \\ r \end{array}
\right) m \ldots [m+(q-r)m] [h(n+1)] \ldots [h(n+1)+(r-1)n] \nn \\
&& =\sum_{r=0}^{q+1} \left[ \left( \begin{array}{c} q \\ r-1 \end{array}
\right) + \left( \begin{array}{c} q \\ r \end{array}
\right) \right] m \ldots [m+(q-r)m] [h(n+1)] \ldots [h(n+1)+(r-1)n] \nn
\\
&& =\sum_{r=0}^{q+1} \left( \begin{array}{c} q+1 \\ r \end{array}
\right) m \ldots [m+(q-r)m] [h(n+1)] \ldots [h(n+1)+(r-1)n] \nn \pkt
\eea
This completes the induction. The second equality of identity
\eqoth{\ref{eq:mir}} follows from
\bea & \left. e^{\lambda \partial_{w}
w^{n+1}} \right|_{w=z}={\displaystyle\sum_{j=0}^{\infty}}
\frac{\lambda^{j}}{j!}
{\displaystyle\prod_{k=0}^{j}} (kn+1) z^{2n} = \left. \frac{\partial
 (e^{\lambda w^{n+1}\partial_{w}}w)}{\partial w}\right|_{w=z} 
\pkt
\eea
We have herewith completed the proof of theorem \refoth{\ref{th:mir}}.
\eprf

\noi Hence \eq{\ref{eq:transf}} simplifies 
to\footnote{Note that $\left( \left.
e^{\lambda \partial_{w} w^{n+1}} \right|_{w=z}
\right)^{h} \neq \left. e^{h \lambda \partial_{w} w^{n+1}}
\right|_{w=z}$\bk .}:
\bea
e^{\lambda L_{n}} \: \Phi_{h}(z) \: 
e^{-\lambda L_{n}} \: = \: 
\left( \left. e^{\lambda \partial_{w} w^{n+1}} \right|_{w=z}
\right)^{h} e^{\lambda z^{n+1}\partial_{z}} \Phi_{h}(z)
\pkt
\eea
We can provide the final step with the following theorem:
\bth \label{th:litth1}
Let $A_{z}$ be a linear differential operator acting on $z$, then
\bea
& e^{A_{z}}f(z)=f(\left. e^{A_{w}} w \right|_{w=z}) \nn
\eea
for any function $f(z)$ which can be expanded in a power series.
\eth
\bprf It is sufficient to show the theorem for functions $f(z)=z^{m}$
for any $m\in \bbbn_{0}$. Since $A_{z}$ is a linear differential
operator, we find:
\bea
&\displaystyle{A_{z}^{n}z^{m}=\sum_{n_{1}+n_{2}+
 \ldots +n_{m}=n}\frac{n!}{n_{1}! \ldots
n_{m}!} (\left. A_{w}^{n_{1}}w\right|_{w=z}) \ldots (\left.
A_{w}^{n_{m}}w\right|_{w=z})} \nn \\
\Rightarrow \bk &\displaystyle{e^{A_{z}}z^{m}=\sum_{n\geq 0}
\frac{1}{n!}A_{z}^{n}z^{m}=\sum_{n_{1},\ldots,n_{m}} \frac{1}{n_{1}! \ldots
n_{m}!} (\left. A_{w}^{n_{1}}w\right|_{w=z}) \ldots (\left.
A_{w}^{n_{m}}w\right|_{w=z})=(\left.
e^{A_{w}}w\right|_{w=z})^{m} \pkt} 
\nn
\eea
\eprf
\noi Using theorem \refoth{\ref{th:litth1}} we obtain for
\eq{\ref{eq:transf}}:
\bea
& e^{\lambda L_{n}} \: \Phi_{h}(z) \:
 e^{-\lambda L_{n}} \: = \:
 \sum_{j=0}^{\infty}\frac{1}{j!}[\lambda L_{n},\Phi_{h}(z)]_{j}
= \left(\left. e^{\lambda \partial_{w}w^{n+1}}
\right|_{w=z}\right)^{h} \Phi_{h} \left( \left. e^{\lambda
w^{n+1}\partial_{w}}w\right|_{w=z}\right)
\, .
\eea
The main result of this subsection is now at hand
after applying \eqs{\ref{eq:id1}}-\eqoth{\ref{eq:id4}}:
\bth \label{th:transf0}
A primary field $\Phi_{h}(z)$ 
transforms for $n\in\bbbz$ according to
\bea
& e^{\lambda L_{n}} \Phi_{h}(z) e^{-\lambda L_{n}} = \frac{1}{(1-n\lambda
z^{n})^{\frac{n+1}{n}h}} \bk \Phi_{h} \left(\frac{z}{(1-n\lambda
z^{n})^{\frac{1}{n}}}\right) & \com \; n \neq 0 \nn \com  \\
& e^{\lambda L_{0}} \Phi_{h}(z) e^{-\lambda L_{0}} = e^{\lambda h} \bk 
\Phi_{h} \left( e^{\lambda}z \right) \pkt \nn 
\eea
\eth

\subsection{Superprimary fields in the $N=1$ case}
The previous section has to be slightly altered in order to apply 
theorem \refoth{\ref{th:transf0}} for $\scn{1}$. 
The operator product of the stress-energy tensor with superprimary fields 
leads to the
following commutation relations where the superprimary fields are taken
at the superpoint $Z=(z,\theta)$:
\bea
[L_{n},\Phi_{h}(z,\theta)] &=& \left(z^{n+1}\partial_{z}+\frac{1}{2}(n+1)z^{n}\theta
\partial_{\theta} +h(n+1)z^{n}\right)\Phi_{h}(z,\theta ) \com \; n\in \bbbz
\com \nn \\
\ [G_{r},\Phi_{h} (z,\theta)] &=& 
\left(z^{r+\half}(\half \partial_{\theta}-\theta
\partial_{z})-h(r+\frac{1}{2})\theta z^{r-\frac{1}{2}}\right) \Phi_{h}(z,\theta)
\com \; r \in \bbbzh \pkt \nn
\eea 
In exactly the same manner as in the Virasoro case we can find:
\be
e^{\lambda L_{n}} \Phi_{h}(z,\theta) e^{-\lambda L_{n}} =
e^{\lambda(z^{n+1}\partial_{z}+\frac{1}{2}z^{n}(n+1)\theta\partial_{\theta}
+h(n+1)z^{n})} \Phi_{h}(z,\theta) 
\pkt
\ee
If we then split $\Phi_{h}(z,\theta)$ into even and odd parts
$\Phi_{h}(z,\theta)=\varphi_{h}(z)+ \theta \psi_{h}(z)$ 
and act on the two parts
separately we notice that both cases come back to \eq{\ref{eq:mir}}.
\bea
e^{\lambda(z^{n+1}\partial_{z}+\frac{1}{2}z^{n}(n+1)\theta\partial_{\theta}
+h(n+1)z^{n})} \varphi_{h}(z) &=& e^{\lambda(z^{n+1}\partial_{z}
+h(n+1)z^{n})} \varphi_{h}(z) \com \nn \\
 e^{\lambda(z^{n+1}\partial_{z}+\frac{1}{2}z^{n}(n+1)\theta\partial_{\theta}
+h(n+1)z^{n})} \theta \psi_{h}(z) &=& \theta  e^{\lambda(z^{n+1}\partial_{z}
+(h+\frac{1}{2})(n+1)z^{n})} \psi_{h}(z) \pkt \nn
\eea
By using theorem \refoth{\ref{th:mir}} we thus obtain:
\bea
e^{\lambda(z^{n+1}\partial_{z}+\frac{1}{2}z^{n}(n+1)\theta\partial_{\theta}
+h(n+1)z^{n})} \varphi_{h}(z) &=& \left( \left. e^{\lambda \partial_{w}
w^{n+1}} \right|_{w=z} \right)^{h} \varphi_{h}\left( \left. e^{\lambda
w^{n+1} \partial_{w}} w\right|_{w=z}\right)
\com \nn \\
e^{\lambda(z^{n+1}\partial_{z}+\frac{1}{2}z^{n}(n+1)\theta\partial_{\theta}
+h(n+1)z^{n})} \theta \psi_{h}(z) &= & \theta \left( \left.
 e^{\lambda \partial_{w}
w^{n+1}} \right|_{w=z} \right)^{h+\frac{1}{2}} \psi_{h}\left( \left. e^{\lambda
w^{n+1} \partial_{w}} w\right|_{w=z}\right) \; .
\nn
\eea
Taking both results together leads to
\bea
e^{\lambda L_{n}} \Phi_{h}(z,\theta) e^{-\lambda L_{n}} \!&=&\! \left(\left.
e^{\lambda \partial_{w} w^{n+1}} \right|_{w=z}\right)^{h}
\Phi_{h}\left(\left. e^{\lambda\partial_{w}w^{n+1}} w\right|_{w=z}, \left(
 \left. e^{\lambda \partial_{w} w^{n+1}} \right|_{w=z}
\right)^{\frac{1}{2}} \theta \right) \, .
\eea
Using \eqs{\ref{eq:id1}}-\eqoth{\ref{eq:id4}} we obtain the
transformation rules for the even generators.
\bth
In the even sector the transformation rules are ($n\in \bbbz$):
\bea
e^{\lambda L_{n}}\Phi_{h}(z,\theta)e^{-\lambda L_{n}}&=&\frac{1}{(1-n\lambda
z^{n})^{\frac{n+1}{n}h}} \Phi_{h}\left(\frac{z}{(1-n\lambda
z^{n})^{\frac{1}{n}}},\frac{\theta}{(1-n\lambda
z^{n})^{\frac{n+1}{2n}}}\right) \com \; n\neq 0 \com \nn \\
e^{\lambda L_{0}}\Phi_{h}(z,\theta)e^{-\lambda L_{0}}&=&e^{\lambda
h}\Phi_{h}\left(e^{\lambda}z,e^{\frac{\lambda}{2}}\theta \right) \pkt \nn 
\eea
\eth
For the odd sector the calculation is less spectacular since the
Taylor expansion comes to an end
after the very first order.
\bth
In the odd sector the transformation rules are ($r\in \bbbzh$):
\bea
e^{\epsilon G_{r}} \Phi_{h}(z,\theta) e^{-\epsilon G_{r}} &=&
[1-\epsilon \theta (r+\half) z^{r-\half}]^{h} \Phi_{h}
\left( z-\epsilon z^{r+\half}\theta ,\theta+\half \epsilon
z^{r+\half} \right) \com \nn 
\eea
where $\epsilon$ is an anticommuting quantity.
\eth
We point out the fact that both results are consistent with the definition
of superprimary fields [\eq{\ref{eq:sprim}}].

\subsection{Superprimary fields in the $N=2$ case}
Thanks to the complex coordinates 
\eqoth{\ref{eq:compl1}}-\eqoth{\ref{eq:compl3}} 
we can write the charged superprimary fields in the $N=2$ case 
as $\Phi_{h,q}(z,\theta^{+},\theta^{-})
=\varphi_{h,q}(z)+\theta^{+}\psi^{-}_{h,q}(z)+\theta^{-}\psi^{+}_{h,q}(z)
+\theta^{+}\theta^{-}\chi_{h,q}(z)$.
As in the previous subsection the commutators 
of $\sct$ elements with superprimary fields
can be written as differential operators acting on the superprimary
fields. These relations are given in the 
\eqs{\ref{eq:phi_cr}}. We then try to find the action of the exponential 
of these differential operators on the 
fields $\varphi_{h,c}(z)$, $\theta^{+}\psi^{-}_{h,q}(z)$, 
$\theta^{-}\psi^{+}_{h,q}(z)$ and $\chi_{h,q}(z)$. 
Once again theorem \refoth{\ref{th:mir}} is 
the main tool in our calculation. After analysing all possible cases
we obtain:
\bea
e^{\lambda L_{n}} 
\Phi_{h,q}(z,\theta^{+},\theta^{-}) e^{-\lambda L_{n}} &=&
\left[ \frac{1}{(1-n\lambda z^{n})^{\frac{n+1}{n}}}
\right]^{h+\frac{nq}{2}
\frac{\theta^{+}\theta^{-}}{z}} \nn \\
&& \Phi_{h,q} \left( \frac{z}{(1-n\lambda z^{n})^{\frac{1}{n}}} ,
\frac{\theta^{+}}{(1-n\lambda z^{n})^{\frac{n+1}{2n}}} ,
\frac{\theta^{-}}{(1-n\lambda z^{n})^{\frac{n+1}{2n}}} \right) 
\label{eq:N2transf1} , \\
e^{\lambda L_{0}} \Phi_{h,q}(z,\theta^{+},\theta^{-}) e^{-\lambda L_{0}}
&=& e^{\lambda h} \Phi_{h,q}\left(e^{\lambda}z,
e^{\frac{\lambda}{2}} \theta^{+}, e^{\frac{\lambda}{2}} \theta^{-}\right) 
\com \\
e^{\epsilon G_{r}^{+}} \Phi_{h,q}(z,\theta^{+},\theta^{-})
e^{-\epsilon G_{r}^{+}} &=&
\frac{1}{[1-\epsilon\theta^{+}
(r+\half)z^{r-\half}]^{2(h+\frac{q}{2})}} \nn \\
&& \Phi_{h,q}\left(
\frac{z}{1-\epsilon\theta^{+}z^{r-\half}},
\theta^{+},\frac{\theta^{-}-\epsilon z^{r+\frac{1}{2}}}{1-
\epsilon\theta^{+}(r+\half)z^{r-\half}}
\right) \com \\
e^{\epsilon G_{-\half}^{+}} \Phi_{h,q}(z,\theta^{+},\theta^{-})
e^{-\epsilon G_{-\half}^{+}} &=&
\Phi_{h,q}(z+\epsilon \theta^{+},\theta^{+},\theta^{-}-\epsilon) \com \\
e^{\epsilon G_{r}^{-}} \Phi_{h,q}(z,\theta^{+},\theta^{-})
e^{-\epsilon G_{r}^{-}} &=& \frac{1}{[1-\epsilon\theta^{-}
(r+\half)z^{r-\half}]^{2(h-\frac{q}{2})}} \nn \\
&& \Phi_{h,q}\left(
\frac{z}{1-\epsilon\theta^{-}z^{r-\half}},
\frac{\theta^{+}-\epsilon z^{r+\frac{1}{2}}}{1-
\epsilon\theta^{-}(r+\half)z^{r-\half}}
,\theta^{-}\right) \com \\
e^{\epsilon G_{-\half}^{-}} \Phi_{h,q}(z,\theta^{+},\theta^{-})
e^{-\epsilon G_{-\half}^{-}} &=&
\Phi_{h,q}(z+\epsilon \theta^{-},\theta^{+}-\epsilon,\theta^{-}) \com \\
e^{\lambda T_{m}} \Phi_{h,q}(z,\theta^{+},\theta^{-}) e^{-\lambda T_{m}}
&=& \! e^{\lambda q z^{m}} \frac{1}{(1-\lambda m\theta^{+}\theta^{-}z^{m-1})^{h}}
\Phi_{h,q}\left(z,e^{-\lambda z^{m}}\theta^{+},e^{\lambda z^{m}} \theta^{-}
\right)  , \label{eq:N2transf2} 
\eea
where $m\in \bbbz$, $n\in \bbbz\without\{0\}$ and $r\in\bbbzh\without\{-
\half\}$. It is easy to check that for $q=0$ we obtain a transformation
according to the definition of (uncharged) superprimary fields 
[\eq{\ref{eq:sprim}}]
where the transformed super Jakobi determinant takes the form
\bea
D\bar{\theta} &=& \left( \begin{array}{cc}
D^{+}\bar{\theta}^{-} & D^{+}\bar{\theta}^{+} \\
D^{-}\bar{\theta}^{-} & D^{-}\bar{\theta}^{+}
\end{array} \right) \pkt
\eea

\subsection{Uncharged superprimary fields}
Similarly to the previous subsections the commutator
\eqoth{\ref{eq:primcom}} allows us to investigate the transformation
properties of uncharged superprimary fields for any parameter $N$. 
The problem comes down to the evaluation of
the action of the differential operator
\bea
{\cal T}_{a}^{i_{1},\ldots,i_{I}} &=&
(-1)^{NI} \Bigl\{ h(a+1-\frac{I}{2})\theta_{i_{1}} \ldots
\theta_{i_{I}} z^{a-\frac{I}{2}} 
+ \theta_{i_{1}} \ldots
\theta_{i_{I}} z^{a+1-\frac{I}{2}} \partial_{z_{2}} \nn \\
&&
-\frac{(-1)^{I}}{2} \sum_{p=1}^{I} (-1)^{p}
\theta_{i_{1}} \ldots \check{\theta}_{i_{p}} \ldots
\theta_{i_{I}} z^{a+1-\frac{I}{2}} \partial_{\theta_{i_{p}}}
\\
&&
+\frac{(a+1-\frac{I}{2}) (-1)^{I}}{2}
\theta_{i_{1}} \ldots 
\theta_{i_{I}} \theta_{j} 
z^{a-\frac{I}{2}} \partial_{\theta_{j}} \Bigr\} \com \nn
\eea
on the superprimary field $\Phi_{h}(Z)$.
The most interesting cases are the Virasoro generators $L_{m}=J_{m}$ and
the generators $G_{r}^{k}=J_{r}^{k}$.
Again theorem \refoth{\ref{th:mir}} is the main device:
\bea
\lefteqn{e^{\lambda L_{n}} \Phi_{h}(z,\theta_{1},\ldots,\theta_{N}) 
e^{-\lambda L_{n}} \:=} \sk \nn \\
&& \frac{1}{(1-n\lambda z^{n})^{\frac{n+1}{n}h}}
\Phi_{h}\Bigl(\frac{z}{(1-n\lambda z^{n})^{\frac{1}{n}}}
, \frac{\theta_{1}}{(1-n\lambda z^{n})^{\frac{n+1}{2n}}}
,\ldots ,\frac{\theta_{N}}{(1-n\lambda z^{n})^{\frac{n+1}{2n}}}\Bigr)
\com \\[1ex]
\lefteqn{e^{\lambda L_{0}} \Phi_{h}(z,\theta_{1},\ldots,\theta_{N}) 
e^{-\lambda L_{0}} \:=\:
e^{\lambda h}\Phi_{h}\left( e^{\lambda h}z,e^{\frac{\lambda}{2}}\theta_{1},
\ldots, e^{\frac{\lambda}{2}}\theta_{N} \right) \com\sk n\in\bbbz
\without \{0\} \com} \sk \\[1ex]
\lefteqn{e^{\epsilon G_{r}^{k}} \Phi_{h}(z,\theta_{1},\ldots,\theta_{N})
e^{-\epsilon G_{r}^{k}} \:=\:
\left(1+(-1)^{N}\epsilon\theta_{k}(r+\half)z^{r-\half}\right)^{h}}\sk\nn\\
&&
\Phi_{h}\Bigl(
z+(-1)^{N}\epsilon\theta_{k}z^{r+\half},
\theta_{1} - \frac{(-1)^{N}}{2}\epsilon\theta_{k}\theta_{1}(r+\half)
z^{r-\half}, \ldots, \nn \\
&&
\theta_{k} - \frac{(-1)^{N}}{2}\epsilon z^{r+\half},
\ldots,
\theta_{N} - \frac{(-1)^{N}}{2}\epsilon\theta_{k}\theta_{N}(r+\half)
z^{r-\half}\Bigr) \com \sk r\in\bbbzh \pkt
\eea
In particular $G_{-\half}^{k}$ is the supertranslation in $\theta_{k}$ 
direction:
\bea
e^{\epsilon G_{-\half}^{k}} \Phi_{h}(z,\theta_{1},\ldots,\theta_{N})
e^{-\epsilon G_{-\half}^{k}} &=&
\Phi_{h}\left(z,
\theta_{1}, \ldots,
\theta_{k} - \frac{(-1)^{N}}{2}\epsilon,
\ldots,
\theta_{N}\right) \com
\eea
and $L_{-1}$ translates in $z$ direction:
\bea
e^{\lambda L_{-1}} \Phi_{h}(z,\theta_{1},\ldots,\theta_{N})
e^{-\lambda L_{-1}} &=&
\Phi_{h}\left(z+\lambda,
\theta_{1}, \ldots,
\theta_{N}\right) \pkt
\eea


\section{{Conclusions and Prospects}}

We presented a superfield framework based on superconformal manifolds
to derive all superconformal Neveu-Schwarz theories. This enabled us
to derive the most general OPE for the stress-energy tensor with
itself and with (uncharged) primary fields for any $N$. 
The resulting commutation relations agree for the classical case $C=0$ with
the commutation relations of superderivatives on the space of
differential forms over a superconformal manifold leaving the
differential form $\kappa$ invariant. This is the exact analogue to
the de Witt algebra being a representation of the Virasoro algebra
for the classical case $C=0$. The general expressions given for the
generators of the Neveu-Schwarz $N$ algebra and their
commutators can be extremely helpful in many applications of conformal
field theory. As an example we computed the transformation 
properties of (uncharged) primary fields under superconformal
transformations generated by algebra generators. 

The representation theories of Neveu-Schwarz algebras 
are known for the Virasoro
case $N=0$ and for $N=1$. However, already for $N=2$ one finds
new structures like degenerated singular vectors\cite{paper1,paper2,beatriz2}
and subsingular vectors\cite{beatriz1,beatriz2,paper7}. 
Besides, the embedding structure
of singular vectors of the $N=2$ Neveu-Schwarz algebra is only known
up to subsingular vectors\cite{thesis}. The derivation of the 
Neveu-Schwarz algebras presented in this paper is valid for 
arbitrary $N$ and therefore serves as a framework for the
study of the representation theory of superconformal field theories
for any $N$. This, however, first requires that
the charged primary fields for arbitrary $N$ would be defined.

Finally, we will present in a forthcoming publication\cite{paper8}
how to derive the Ramond superconformal field theories using a similar 
superfield formalism.


\appendix

\acknowledgements


\noi
I am very grateful to Adrian Kent for numerous illuminating
discussions and to Victor Kac for many important comments.
I would like to thank Matthias Gaberdiel and 
G\'erard Watts for various helpful remarks. 
Special thanks go to 
my wife Val\'erie for her support in 
many linguistic matters. This work was supported by a  
DAAD fellowship, by SERC and in part by NSF grant PHY-92-18167.


\newcommand{\tit}[1]{{\it #1}}

\end{document}